\documentclass[manuscript]{acmsmall}

\usepackage{amssymb,amsmath}
\usepackage{graphicx}
\usepackage{url}
\usepackage{caption}
\usepackage{amsmath,amsfonts}
\usepackage[ruled]{algorithm2e}

\SetAlFnt{\small}
\SetAlCapFnt{\small}
\SetAlCapNameFnt{\small}
\SetAlCapHSkip{0pt}
\IncMargin{-\parindent}

\usepackage{xcolor}

\usepackage[normalem]{ulem}

\begin{document}

\title{Reliability or Sustainability: Optimal Data Stream Estimation and  Scheduling in Smart Water Networks}

\author
{
	SOKRATIS KARTAKIS
	\affil{Imperial College London}
	SHUSEN YANG
	\affil{Xi'an Jiaotong Univerity}
	JULIE A. McCANN
	\affil{Imperial College London}
}

\begin{abstract}

As a typical Cyber-Physical System (CPS), smart water distribution networks require monitoring of underground water pipes with high sample rates for precise data analysis and  water network control. Due to poor underground wireless channel quality and long-range communication requirements, high transmission power is typically adopted to communicate high-speed sensor data streams; posing challenges for long term sustainable monitoring. In this paper, we develop the first sustainable water sensing system, exploiting energy harvesting opportunities from water flows. Our system does this by scheduling the transmission of a subset of the data streams, while other correlated streams are estimated using auto-regressive models based on the sound-velocity propagation of pressure signals inside water networks. To compute the optimal scheduling policy, we formalize a stochastic optimization problem to maximize the estimation reliability, while ensuring the system’s sustainable operation under dynamic conditions. We develop Data Transmission Scheduling (DTS), an asymptotically optimal scheme; and FAST-DTS, a lightweight online algorithm that can adapt to arbitrary energy and correlation dynamics. Using over 170 days of real data from our smart water system deployment and conducting in-vitro experiments to our small-scale testbed; our evaluation demonstrates that Fast-DTS significantly outperforms three alternatives, considering data reliability, energy utilization, and sustainable operation.
\end{abstract}


\maketitle

\section{Introduction}
Optimal water distribution and energy waste reduction are currently hot topics. Water demands are not being met in many regions around the globe; both developed and underdeveloped; where climate change and economic water scarcity are two issues that have the largest impact. Both drought prone and wet areas have observed severe water network operation problems that have lead to water restrictions and losses respectively. Notwithstanding the 7.5bn investment in UK water distribution networks, 3.3bn liters of water were lost per day in 2010 \cite{johnson2010water}.

Over the last decade, there has been a trend for water utility companies to create smart water networks in order to improve the quality of service, reduce water waste through balancing the water supply and demand, and minimize the maintenance cost by increasing network resilience. To achieve these goals, wireless sensing technologies are being adopted for the monitoring of water network states and the detection of abnormal behaviors such as water leakage and bursts \cite{narayanan2014one,aghaei2011using,santos2011sensor,zhu2010remote,lijuan2012leak,kartakis2016adaptive}, which feeds the precise control of the network. Therefore, the reliability of sensing and anomaly detection plays a key role in the success of such networked Cyber-Physical Systems (CPSs) as a whole.

According to the reports from water utility companies \cite{swig2014}, the main limitations of current water distribution network infrastructures are: (a) the underground position of water network assets, such as Water Meters, Pressure Reducing Valve (PVR) Controllers, Pressure Transducers, Acoustic Leak Detectors, and District Metering Area (DMA) Meters, (b) the country-scale deployment, and (c) the lack of power to these underground locations. Specifically, less than 0.5\% of the assets are above ground, more than 99\% are remote away from power, and the water networks often flow in geographically remote un-populated areas. Therefore, the provision or maintenance of power and wired communication within underground asset locations is generally considered too costly and the battery-driven wireless sensor nodes are really the only real choice. For the same reasons, wide spread wired telemetry and control systems which deployed in industrial environments, such as Supervisory Control and Data Acquisition (SCADA) systems \cite{scada2009resource}, to support city-scale infrastructures incorporate both radio and wired communication.

In smart water networks, physical states such as water pressure and pipe vibration need to be sampled at a high frequencies (e.g. more than 128 samples per second per sensor). This high frequency is required to capture instabilities or transient event of fast signals, like pressure signals propagated with sound velocity inside the pipes. Additionally, high sample-rated data can be exploited by server-side algorithms to generate high level of information precisely, such as the burst localization algorithm in [1], which localize bursts with 0.5m accuracy.

However, to transmit these required high-speed raw sensor data streams through long-range (several kilometers) wireless communications, high transmission power is required that lead to fast battery depletion. As a result, utility companies who have already spent billions of dollars in water network maintenance, has to provide the supplementary expensive maintenance for frequently replacing batteries, is considered unaffordable. Therefore, sustainable smart water sensing system design for autonomous water network monitoring, which balances the communication, is highly desired.

Current smart water systems \cite{allen2013water,stoianov2007pipenet} are unable to achieve the above objectives in a cost-effective way. For instance, the sustainable smart water system proposed by MIT~\cite{allen2013water} adopts large overground solar panels and direct power from lampstands, requiring expensive instrumentation and deployment costs; while other approaches such as \cite{lijuan2012leak} require knowledge of the complete hydraulic models of water networks, which is computationally expensive and unable to adapt to water network system dynamics.

\subsection{Our Approach}

\begin{figure}
	\centerline{\includegraphics[width=3.9in]{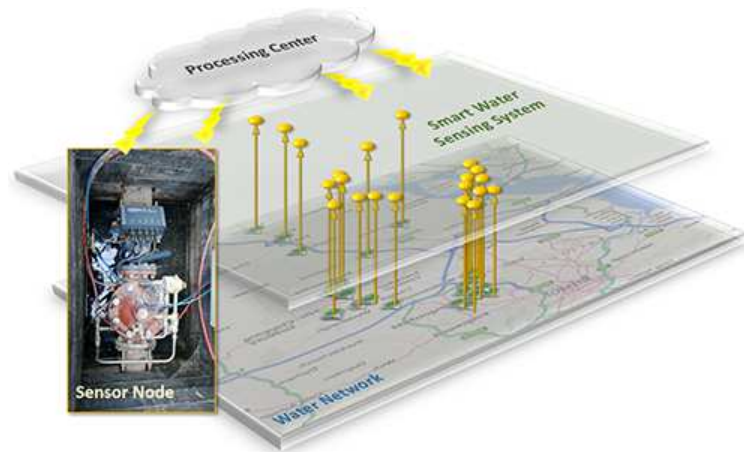}}
	\caption{An illustration of our smart water system and sensor node hardware.}
	\label{fig:Fig1}
\end{figure}

In this paper, we propose a cost-effective and sustainable sensing system for the autonomous monitoring of system states and abnormal behaviors in water networks. Contemporary water supply network structure consists of three individual layers: (a) storage and pumping, (b) supply zones and District Meter Areas (DMAs), and (c) end users (customers). While valves control flows and pressures at fixed points in the water network, pumps pressurize water to overcome gravity and frictional losses along supply zones, which are divided into smaller fixed network topologies (in average 1500 customer connections) with permanent boundaries, DMAs. The water pressure distribution and flows into each DMA is continuously monitored with the aim to enable proactive leakage management, simplistic pressure management, and efficient network maintenance.

In our project, 24 sensor/actuator nodes were deployed to the inlet of equal number of DMAs in Bristol Water network, which record, analyze, and transmit high sample-rate pressure and flow data (up to 128 Samples/sec) to a data processing center periodically, and cover an area of approximately 7,544 customer connections and 57km of pipe mains as shown in Fig.~\ref{fig:Fig1}. A low-frequency (868 MHz) and high-power wireless communications was utilized (e.g. around several Millijoules per Kilobytes \cite{digi2009xbee868}) to support the long-range and high-speed sensor data stream transmissions.

\begin{figure}
	\centerline{\includegraphics[width=4in]{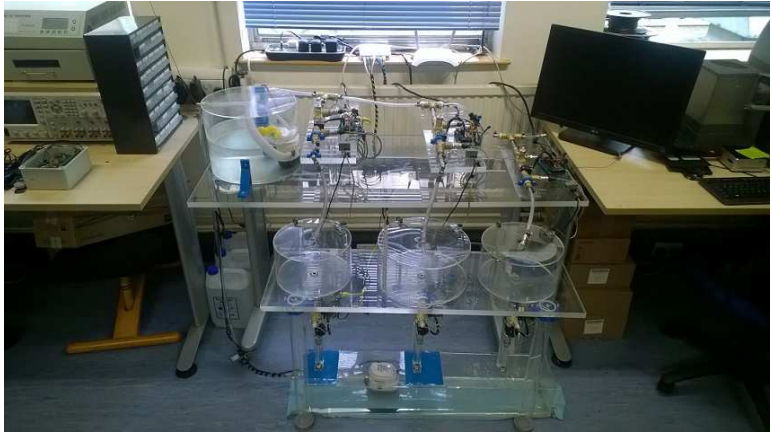}}
	\caption{WaterBox: a small-scale testbed.}
	\label{fig:Fig1w}
\end{figure}

Additional capability of our large scale smart water distribution networks (WDN) is the remote control of DMA actuator components (i.e. valves) which optimize the water network performance and lifetime over varying demands. Because of underground power limitations, main trend is the development and deployment of self-powered actuators. In our system, self-powered multifunction network actuators for dynamically reconfigurable DMAs have been developed which integrate a Cal-Val 99-51 \cite{claval2014valve} and an energy
generator that harvests energy (up to 0.7 W) from the water flow \cite{claval2014}. Under the context of the same project, an in-vitro small scale testbed was created to emulate a smart water network, which consist of three DMAs, and evaluate the efficiency of our algorithms, WaterBox \cite{kartakis2015waterbox} (Fig.~\ref{fig:Fig1w}).

In order to achieve sustainable sensor node operation by utilizing the limited and dynamic harvested energy, we use lossless compression and propose an \textit{in-node} anomaly detection algorithm to reduce the raw data volumes (and therefore the energy) required for the high-power transmission. To further reduce energy consumption,  our system only requires to transmit a subset of the raw data streams, while other samples are estimated using auto-regressive models \cite{papadimitriou2013dimensionality}, based on strong correlations among sensor data streams observed from real data. Fundamentally, these correlations are based on the sound-velocity propagation of pressure signals inside the water network.

In recent years, advanced control methods have been considered in pressure management and pump scheduling tasks (i.e \cite{lopez2008ant}); all these require reliable information about the system states. However, this information is not always available, due to sensors' hardware malfunction or battery depletion, and data estimation is required. As a result, the unavoidable estimation error not only leads to unreliable sensor data analytics, but also the performance degradation of smart water control system as a whole. Therefore, this paper aims to maximize the accuracy of all estimated data streams by fully exploiting the dynamic energy-harvesting opportunities.

The operation of our system is as follows: initially the data processing center receives raw data and energy information from all sensor nodes in the network. After this, the data processing center uses the raw sensor data to establish a correlation graph of the network. This and the energy information are used to determine which nodes are now required to send raw data; this instruction is sent individually to those nodes only. For every subsequent time interval (we set this to 15 minutes in our system) all nodes are required to make note of their behaviours using anomaly detection results and energy metadata to the server. Therefore, unless the data processing center requests it, only the anomaly results and energy information are communicated to the server for each interval.

It can be seen that a fundamental question in our system is: \textit{which are the best subsets of sensor nodes that should be requested to transmit their raw data, given the complex system dynamics regarding the (arbitrary) stochastic processes of energy harvesting, energy consumption, and correlations among sensor data streams?} We answer this question by formalizing a stochastic optimization problem to maximize the estimation reliability while ensuring the sustainable operation of systems, and solve the problem by developing a lightweight algorithm with strong theoretical guarantees to compute the best transmission nodes at real-time.

\subsection{Contributions}
The specific contributions of this paper are as follows:
\begin{itemize}
\item To reduce the energy consumption for long-range high-power data transmissions, we develop an new \textit{in-node} anomaly-detection algorithm that identifies the abnormal behaviours (e.g. bursts and leakages) in water networks and  propose an estimation-based transmission solution by exploiting the properties of sound-velocity propagation of pressure signals inside the water network. Different from current anomaly detection algorithms, we adopt an novel approach that detects anomalies by analyzing data compression rates rather than raw data. In addition, our estimation is based on lightweight auto-regressive models, avoiding the use of complex hydraulic models\cite{lijuan2012leak}.
 \item We formalize a stochastic optimization problem for the best selection of raw data transmissions, which aims to maximize the aggregated estimation reliabilities while guarantee a minimal reliability constraint and the sustainable operation of the smart water sensing system.
\item We develop Data Transmission Scheduling (DTS), an asymptotically optimal solution to the formalized problem, based on Lyapunov optimization theory \cite{neely2010stochastic}. Guided by the principles of DTS, we then propose FAST-DTS, a lightweight online algorithm that can adapt to Both DTS and FAST-DTS. Both DTS and FAST-DTS \textit{do not} need to predict any future knowledge of the water network. Further, we \textit{do not} make any stochastic/probabilistic assumptions regarding the system dynamics, which means that our approach is adaptive to arbitrary energy and correlation dynamics.
\item Our work bridges the gap between data processing and resource allocation of wireless systems. To our knowledge, this is not only the \textit{first} approach that adopts data stream estimation in smart water systems, but also the \textit{first} scheduling approach based on data stream estimation in energy harvesting networks \cite{sudevalayam2011energy}.
\end{itemize}

We evaluated our system by using 170-day water pressure data from our real smart water system. During this process, we examined data reliability, energy waste, node lifetime, and transmission gaps by using three different algorithms where FAST-DTS outperformed them. Further evaluation was conducted by using and extending the hardware infrastructure of the small scale testbed, WaterBox \cite{kartakis2015waterbox}, to verify the adaptive behavior of the system in anomalies (i.e. leakages and bursts).

\subsection{Related Work}
Many current research on smart water networks \cite{allen2013water,stoianov2007pipenet} focus on efficient sensing system design , while others \cite{narayanan2014one,aghaei2011using,santos2011sensor,zhu2010remote,lijuan2012leak} seek leakage detection solutions by developing new anomaly detection algorithms.
 However, none of them considers energy optimization nor data stream estimation.
 Similar to our work, \cite{allen2013water} considers energy harvesting systems for water monitoring, but this work require expensive instrumentation and deployments of solar panels and lampposts (using solar power at daytime, and lampposts at night). In contrast, our system harvests energy from water flows, resulting easy implementation, and continuous power supply (both day and night). More importantly, our scheduling algorithm (including power management) can ensure the sustainable operation of the system. In fact, there is an increasing interest in developing more efficient water flow energy harvester~\cite{pobering2008power,hoffmann2013energy,morais2008sun}.
In addition, optimization for energy harvesting networks is an emerging hot research area \cite{sudevalayam2011energy,harvestingdelay,huang2013utility,yang2013distributed,liu2011perpetual,liu2010joint}. A fundamental research issue for these approaches are to optimize the network performance (e.g. delay~\cite{harvestingdelay} and network utility~\cite{huang2013utility,liu2010joint}) while ensuring the sustainable operation of the network, by developing network algorithms such as routing and flow rate control. However, none of them considers smart water networks, nor the problem of estimation reliability maximization that we study in this paper.

Steady state water distribution networks can be considered similar to industrial process plants such as precise industrial temperature control systems. However the combination of large-scale, dynamic reconfiguration due to anomalies (i.e. bursts, leakages), and nondeterministic behavior of water networks due to demand change along the time, set this comparison unreal and impractical. These challenges introduce new needs to the control process, which are unnecessary for industrial process plants which are isolated systems. Automatic control methods are vitally important in solving some operational challenges like reducing pressure driven leakage, energy usage in pumping, leak localization etc. However, a naive deployment of new control technologies in critical infrastructure and their potential failures may have catastrophic consequences for large-scale operational smart water networks. Two of the few sophisticated operational control systems include i2O \cite{i2o2015} and Derceto \cite{derceto2015}, who have developed and applied 'optimal' automatic remote control of PRVs and pumps, respectively. However, these systems are based on diurnal training and communication that produces static control schedules for predefined periods of time.
In our project, we study and develop real-time control algorithms, which require frequent communication with sensor/actuator nodes to achieve optimal network reconfiguration in anomalies and demand changes. Thus, the development of a sustainable smart water network is essential.

\subsection{Paper Organization}
The rest of this paper is organized as follows. Section \ref{sec:preliminary} briefly provides preliminaries for the understanding of water pressure signal transmission and data stream correlations. Section \ref{sec:system} describes the system overview and models. Section \ref{sec:anomaly} presents the anomaly detection algorithm. Section \ref{sec:Scheduling} presents the DTS and FAST-DTS algorithms. Section \ref{sec:estimation} discusses auto-regressive models and the estimation process of water pressure data. Section \ref{sec:evaluation} presents our evaluation, an we finally conclude this paper in Section \ref{sec:conclusion}.

\section{Preliminary}
\label{sec:preliminary}
Pressure waves are generated at any point in a pipe system where a disturbance occurs because of flow rate change. An external disturbance could be a valve that is opening or closing, a pump that is started up or shut down, a change in reservoir pressure, or in the inflow or outflow for the system. A pressure wave, which represents a rapid pressure and associated flow change, travels at sonic velocity \cite{wood2005numerical} by using the in pipe liquid as a medium, and the wave is partially transmitted and reflected at all discontinuities in the pipe system (pipe junctions, pumps, open or closed ends, surge tanks, etc.).

\begin{figure}
\centerline{\includegraphics[width=4.3in]{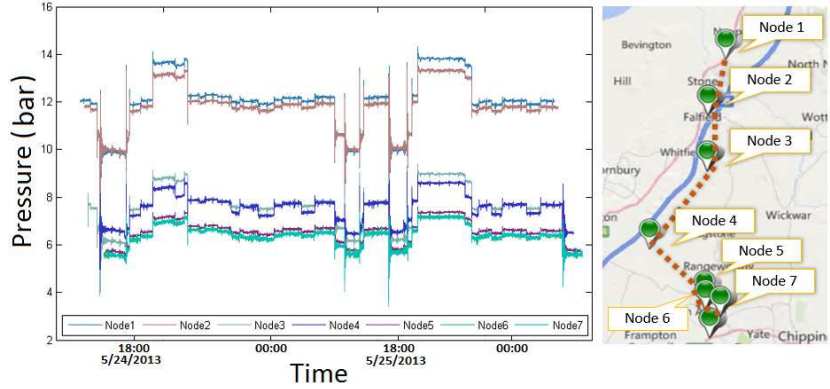}}
\caption{Pressure wave propagation delay and attenuation of sequential nodes}
\label{fig:Fig2}
\end{figure}

The sonic speed, and consequently the delay of the pressure wave propagation, for a liquid flowing within a line are influenced by the elasticity of the line wall, pipe diameter and thickness, and restraint on longitudinal pipe movement. In addition to wave propagation delay, the effect of in-pipe friction causes attenuation in pressure wave, which is related to the length and material of the pipe. Specifically, the equation that describes the pressure wave attenuation/ amplification between two individual points in terms of pressure head based on \cite{wood2005pressure} is the following:
\begin{equation}
H_2 - H_1=\frac{-fLQ^2}{2gDA_L^2}
\label{eq:equationl}
\end{equation}
Where $Q$ is the water flow rate, $L$ is the pipe length, $f$ is the friction factor, $D$ is the diameter, and $A_L$ is the area of the pipe between the two points.
By using well-studied equations like \eqref{eq:equationl}, the propagation delay and wave attenuation in a water network can been modeled accurately. The modeling normally involves the physical properties of different pipes and network topology and this would allow the estimation of the pressure of each interconnected point in a water network using raw data only from one point. However, the modeling task is unfeasible because of the size of a real water network and impossible because of unpredictable dynamic changes (e.g. new asset installation and network expansion) or anomalies (leakages and bursts) in the water network. For these reasons, water network monitoring systems have required the installation of battery driven sensor nodes (of which 99\% are underground) on the main pipes at least, junctions and valves.

One of water network analysts' main interests lie in this system assisting them to predict leakages and network problems. This is focused on the observation of pressure wave behavior and transformations at discontinuities; so called transient events. Because of the high velocity, the analysis of pressure waves requires relatively high sample-rate pressure data from the water network. The initial system that we have installed has the ability to retrieve at least 128 pressure samples per second from DMA inlets across the water network. Figure~\ref{fig:Fig2} represents high sample-rate pressure data from seven sequential nodes, which allows transient event analysis and the observation of propagation delay and the signal attenuation.

\section{System Overview and Modeling}
\label{sec:system}

\begin{figure}
	\centerline{\includegraphics[width=3.5in]{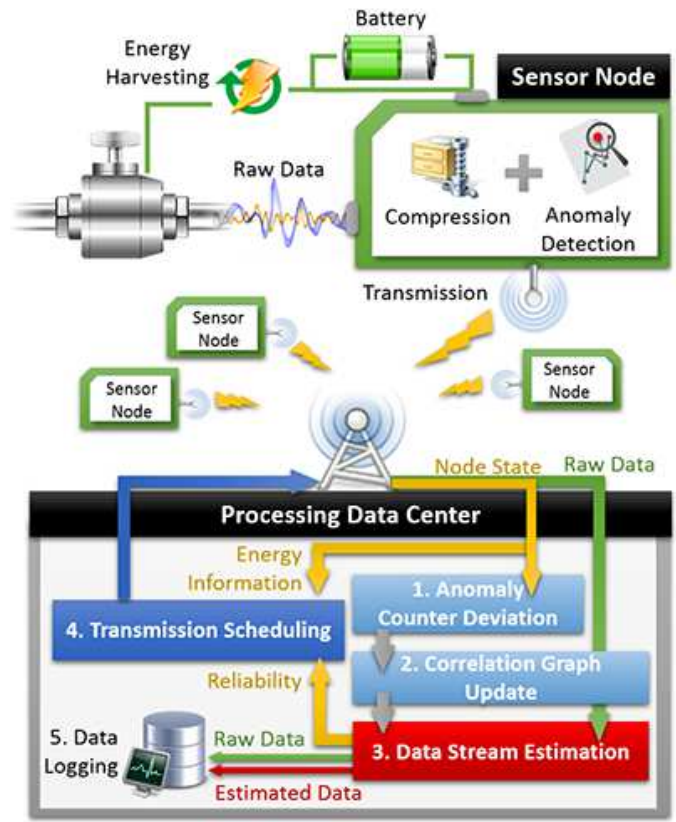}}
	\caption{Illustration of the System architecture.}
	\label{fig:Fig3}
\end{figure}

As shown in Fig.~\ref{fig:Fig3}, a smart water sensing system consists of a set of sensor nodes $\mathcal{S}$ and a data processing center. Continuous physical time is divided into discrete intervals $t=\{1,~2,...,~{\rm t_{end}}\}$ (e.g. a default interval is 15 minutes in our system), where the time-horizon of the smart water system ${\rm t_{end}}$ can be any large but finite value (e.g. several years).

\subsection{Sensor Nodes} At each interval $t$, each sensor node carries out the following operations:

\begin{figure}
	\centerline{\includegraphics[width=3.3in]{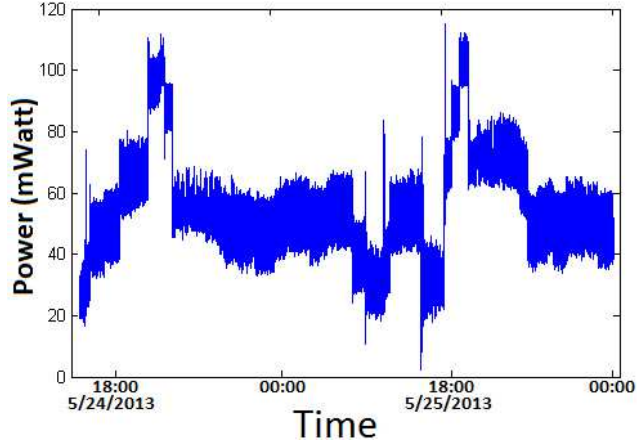}}
	\caption{Dynamic energy generated by water flow.}
	\label{fig:energyharvesting}
\end{figure}

\subsubsection{Sensing and Compression}

Each sensor node collects water pressure data at a high sample rate (e.g. more than 128 samples per second in our system), and formats this data in \textit{chunks} of multiple measurements (e.g. 100 measurements per chunk in our system). A time stamp of a chunk is defined as that of the first measurement in this chunk.
The formatted data is compressed chunk by chunk, using a lightweight lossless compression algorithm (e.g. S-LZW-MC \cite{sadler2006data}) and miniLZO \cite{kraus2008optimal}) which significantly reduces the energy consumption for the high-power wireless transmissions.

\subsubsection{Anomaly Detection}
We develop an efficient algorithm that detects and water network anomalies such as leakage and burst in real-time. The output of the anomaly detection algorithms are the timestamps of anomalies during each interval. In addition, we observe from real data and verified with experiments in the small scale testbed that anomalous behaviors also will have a significant impact on the correlation strength among data streams, which further affects estimation reliability. Therefore, the combination of data estimation and anomaly detection are central for our overall system design. We will discuss the anomaly detection algorithm in detail in Section \ref{sec:anomaly}.

\subsubsection{Wireless Communication}
Water networks cover city scale areas and most of the assets, i.e. sensor nodes, are deployed underground. As a result, long range wireless communication technologies are unavoidable either to cover long distances or to increase signal penetration. For this reasons, the contemporary Low Power Wide Area (LPWA) technologies \cite{moyer2015lpwa}, which are mostly single hop, can be considered as the most appropriate communication approaches. The current paper considers a single hop communication infrastructure based on LPWA technologies and uses the experimental results of \cite{kartakis2016demystifying} to setup the evaluation parameter, such as communication energy consumption. Specifically, the parameters of XBee868 \cite{digi2009xbee868} were used which implements the Zigbee protocol for long ranges over 868MHz.

At this point, the timing limitations of LPWA technologies are important to be mentioned. Based on experimental result in [3], the tradeoff of implementing a reliable long range communication based on low frequency, i.e. 868 MHz, is the decrease of data rate. In our system, every sensor node transmits 60 kBytes of water raw data to a data processing center, while the data rate of XBee868 is 24kbps. Thus, the required time is around 15sec. By the fact that data processing centers are highly capable hardware infrastructure, the required time for decision making processes (i.e. scheduling) is at least one to two orders of magnitude less that the data transmission.

At each interval, each sensor node is required to transmit its  \textit{\textbf{Node States}} information to the data processing center, which includes (a) the number of anomalies within this particular interval denoted as the  \textit{\textbf{Anomaly Counter}}; (b) its energy related parameters which will be discussed later soon.

According to its transmission information, each sensor node $i\in\mathcal{S}$ can provide two communication states at each interval $t$: either: (a) transmit both raw data stream and node states, or (b) transmit node states only. We use a binary variable $y_i(t)=1$ and $y_i(t)=0$ to denote these respectively.

\subsubsection{Power Management}

Each sensor node $i$ can harvest $h_i(t)$ amount of energy from the water flow \cite{claval2014} at each interval $t$. As shown in Fig.\ref{fig:energyharvesting}, $h_i(t)$ is time-varying due to the dynamic water flow rate in the water network.
The harvested energy can either be consumed by in-node operations (i.e. sensing and computation) and wireless transmission; or be stored in the battery, which is modelled as an energy queue
\begin{equation}
0\leq B_i(t)\leq {\rm B_{max}}\label{eq:batterysize}
\end{equation}
where ${\rm B_{max}}$ is the battery capacity. In our system, $\rm B_{max}=60.2$  KJ (two 3.6V batteries with capacities of 2.4Ah each).

Considering the communication states $y_i(t)$ of each node $i$, the queueing dynamic of the battery at each sensor node $i\in \mathcal{S}$ is modelled as

\begin{equation}
B_i(t+1)=|B_i(t)-y_{i}(t)E^{tr}_i(t)- E^{in}_i(t)|_{+}+h_i(t)\label{eq:energybufferupdate}
\end{equation}

where for any real number $x$, $ |x|_{+}=0$ if $x\leq0$, $ |x|_{+}=x$ otherwise. $E^{tr}_i(t)$ and $ E^{in}_i(t)$ represent the energy costs for transmission and all in-node operations (including sensing, compression, and anomaly detection), which are expected to change over time due to the system dynamics such as time-varying channel quality etc. It is worth noting that since the energy cost for wireless transmission in state $y_i(t)=0$ is minimal (its several bytes for node states transmission), we only consider energy consumption caused by in-node operations in this state.

\subsection{Data Processing Center}
Based on the node states and raw data streams received from the sensor nodes, the data processing center is responsible for performing the tasks of correlation graph updates, data stream estimation, and raw data transmission scheduling.

\begin{table}
\centering 
\resizebox{10cm}{!}
{
\begin{tabular}{ |l|l| }
\hline
$\rm t_{end}$ & The finite time-horizon of the system.\\
\hline
$\mathcal{S}$ & The set of all sensor nodes\\
\hline
$G(\mathcal{S},\mathcal{L}(t))$ & The time-varying correlation graph\\
\hline
$\mathcal{N}_i(t)$ & Correlation neighbor set of sensor $i$ at interval $t$\\
\hline
$\mathcal{S}_a(t)$ & Enforced transmission sensor set at interval $t$ \\
\hline
$\mathcal{S}_b(t)$ & Non-enforced transmission sensor set at interval $t$ \\
\hline
$y_i(t)$ & Sensor $i$ 's raw data transmission state at interval $t$ \\
\hline
$\mathcal{Y}(t)$ & Scheduling decision at interval $t$ \\
\hline
$rlb_i(\mathcal{Y}(t))$ & Data estimation reliability of sensor $i$ at interval $t$ \\
\hline
$\rm rlb_{min}$ & minimal estimation reliability requirement \\
\hline
$B_i(t)$ &Battery level of sensor $i$ at interval $t$ \\
\hline
$\rm B_{max}$ & Battery capacity \\
\hline
$h_i(t)$ & Harvested energy by sensor $i$ at interval $t$\\
\hline
$E^{tr}_i(t)$ & Sensor $i$'s transmission energy cost at interval $t$ \\
\hline
$E^{in}_i(t)$ & Sensor $i$'s in-node operation energy cost at $t$ \\
\hline
$\rm B_{exp}$ & Parameter for sustainability/reliability tradeoff \\
\hline
$V$ & Parameter for sustainability/reliability tradeoff \\
\hline
\end{tabular}
}
\label{table:symbol} 
\caption{Frequently used symbols} 
\end{table}

\subsubsection{Correlation Graph Updates}
Due to the water network interconnections, neighboring sensor nodes provide high correlated data (i.e. in Fig.~\ref{fig:Fig2}). The data correlation among different nodes may change either for predictable or unexpected reasons within a water network. Changes to the water network topology because of the control of water network assets, i.e. valves, or the physical changes from engineers, such as isolation of pipes or areas for maintenance, can be considered predictable. Due to the slow physics (e.g. demand changes) of water networks, these actions can occur in the range of some hours to days resulting the rare correlation graph update. On the other hand, anomalies, such as leakages or bursts are unexpected. By exploiting, the anomaly detection at the edge and the a-prior knowledge of water network topology changes, the proposed system can minimize the correlation graph updates accordingly.

\begin{figure}
\centering
	\begin{minipage}[b]{0.45\textwidth}
		\centerline{\includegraphics[width=2.5in]{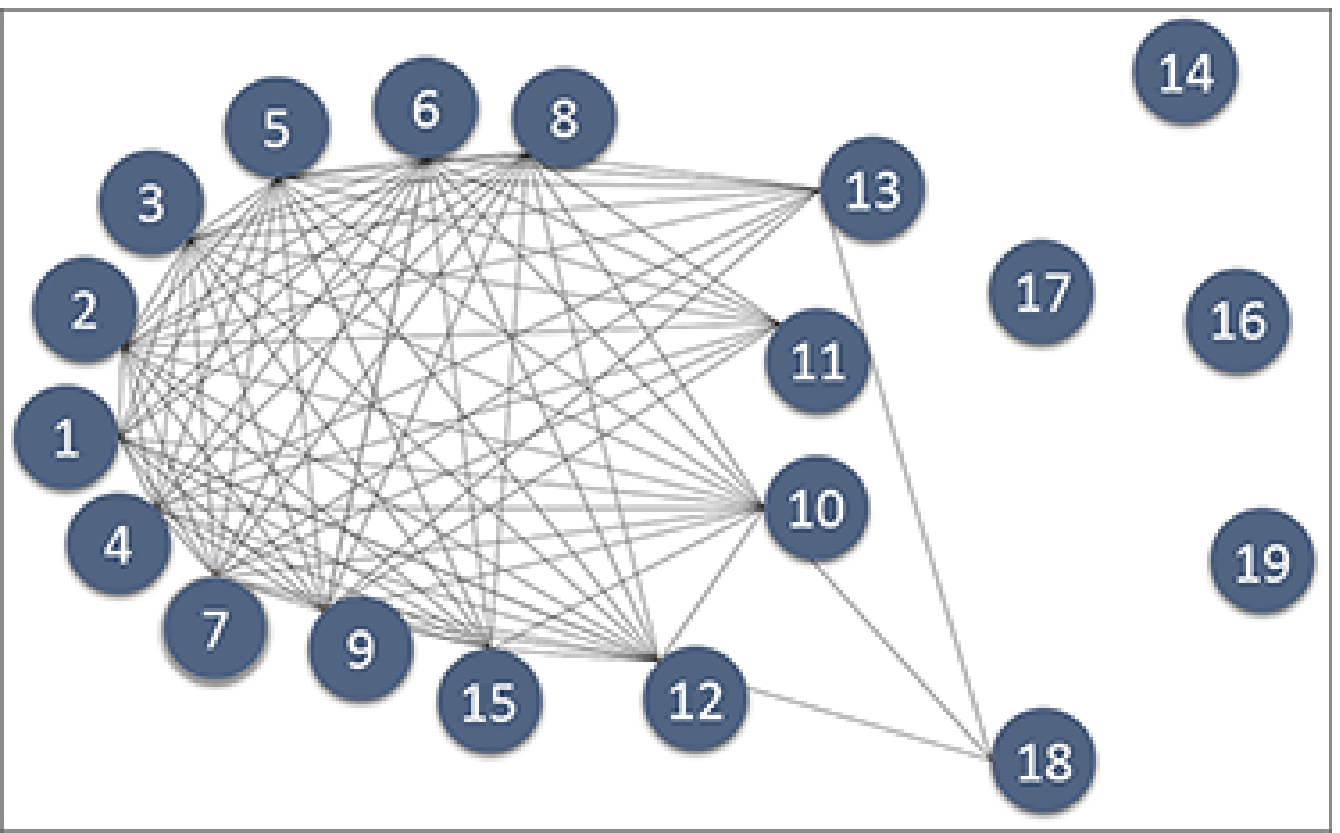}}
		\subcaption{Interval $t_1$}
	\end{minipage}
	\hspace{4mm}
	\begin{minipage}[b]{0.45\textwidth}
		\centerline{\includegraphics[width=2.5in]{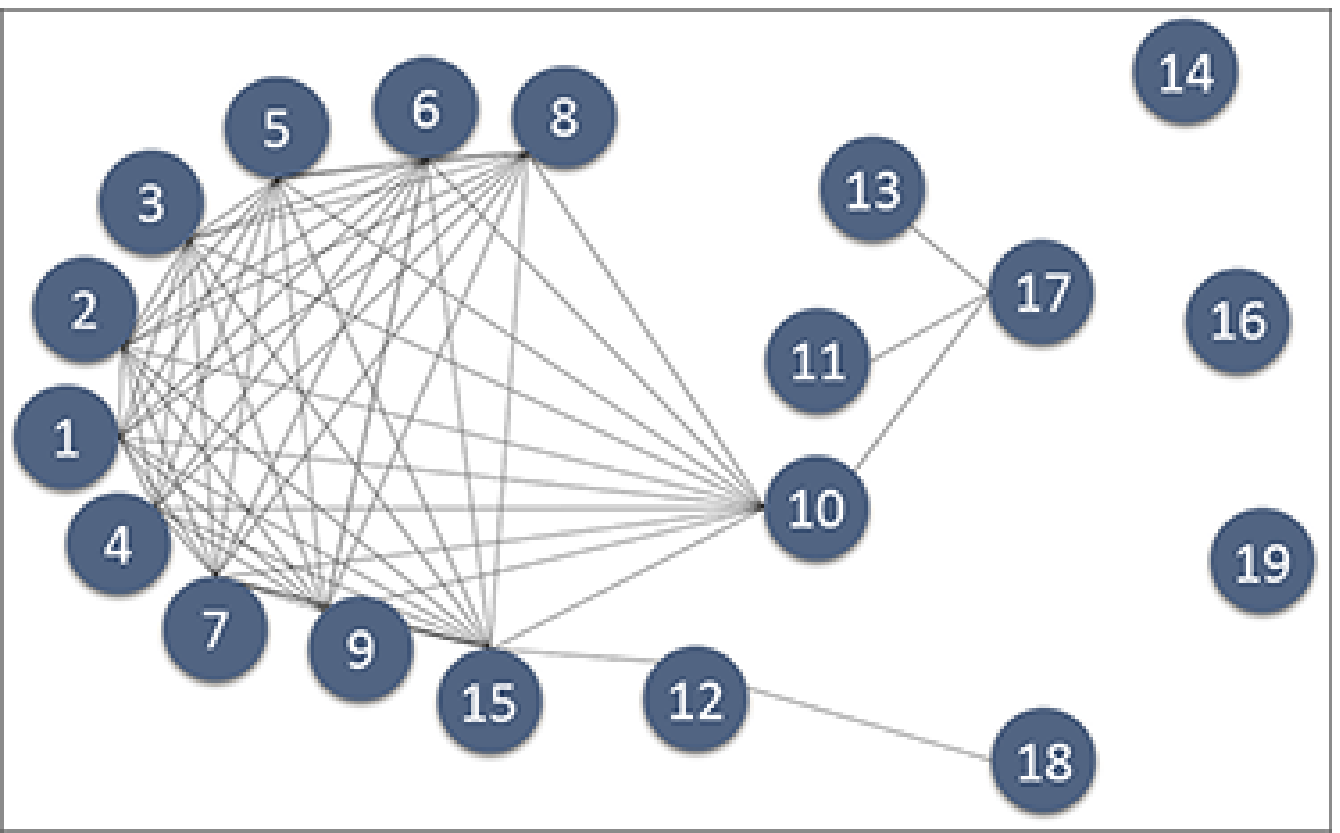}}
		\subcaption{Interval $t_2$}
	\end{minipage}
	\caption{The time-varying correlation graph topology at two intervals $t_1$ and $t_2$.}
	\label{fig:correlationgraph}
\end{figure}

Specifically, the data processing center computes the time-varying correlation graph $G(\mathcal{S}, \mathcal{L}(t))$, where $\mathcal{L}(t)$ represents the set of all current correlation links at each interval $t$. A correlation link $(i,j)$ is considered to exist in $\mathcal{L}(t)$, if the
current Pearson correlation coefficient $c_{i,j}(t)$ of the raw data streams of sensor nodes $i$ and $j$ are larger than 95\%\footnote{95\% is chosen heuristically based on the reliability results of the estimation models in our system.}, i.e.
 \begin{equation}
 \mathcal{L}(t):=\{(i,j):c_{i,j}(t)\geq95\%, i,j\in\mathcal{S}\}
 \end{equation}
 An example of a time-varying correlation graph of our sensor nodes is illustrated in Fig.\ref{fig:correlationgraph}.
For any given sensor $i\in\mathcal{S}$, we define its temporary correlation neighbor set as
 \begin{equation}
 \mathcal{N}_i(t):=\{j:j\in\mathcal{S}, (i,j)\in\mathcal{L}(t)\}\label{eq:neighbor}
 \end{equation}
 i.e. the sets of sensors that currently correlated with $i$.

 The data processing center updates the correlation graph in real-time, \textit{whenever a fresh raw data stream is received.}

\subsubsection{Anomaly Counter Outlier Processing}

In smart water systems, two streams of pressure data are strongly correlated if  their anomaly patterns are similar, observed from real pressure data streams. This is because any change of pressure (e.g. due to the change of the pressure regulator state) is diffused along the water network evenly. In the case when a burst or leakage (or other types of anomalies) occurs, a subpart of the water network is influenced and consequently its correlation. We use the anomaly accouter of each node at every interval as an index for anomaly patterns, to detect the sudden  change of the correlation graph.

After receiving the anomaly counter values of all sensor node for an interval, the data processing center will compute if a high deviation (outlier) of anomaly counters occurs at each sensor node, based on Chauvenet's criterion \cite{sathe2013survey}. If an outlier occurs at a sensor node, say $i$, the correlation pattern between this node and all its current correlation neighbors in $\mathcal{N}_i(t)$ may be changed. In this case, the data stream of $i$ cannot be accurately estimated and the correlation graph should be updated as soon as possible. Therefore, the system will \textbf{enforce} sensor $i$ to transmit its raw data stream for this interval to update the correlation graph.

\begin{figure}
\centering
	\begin{minipage}[b]{0.45\textwidth}
		\centerline{\includegraphics[width=2.9in]{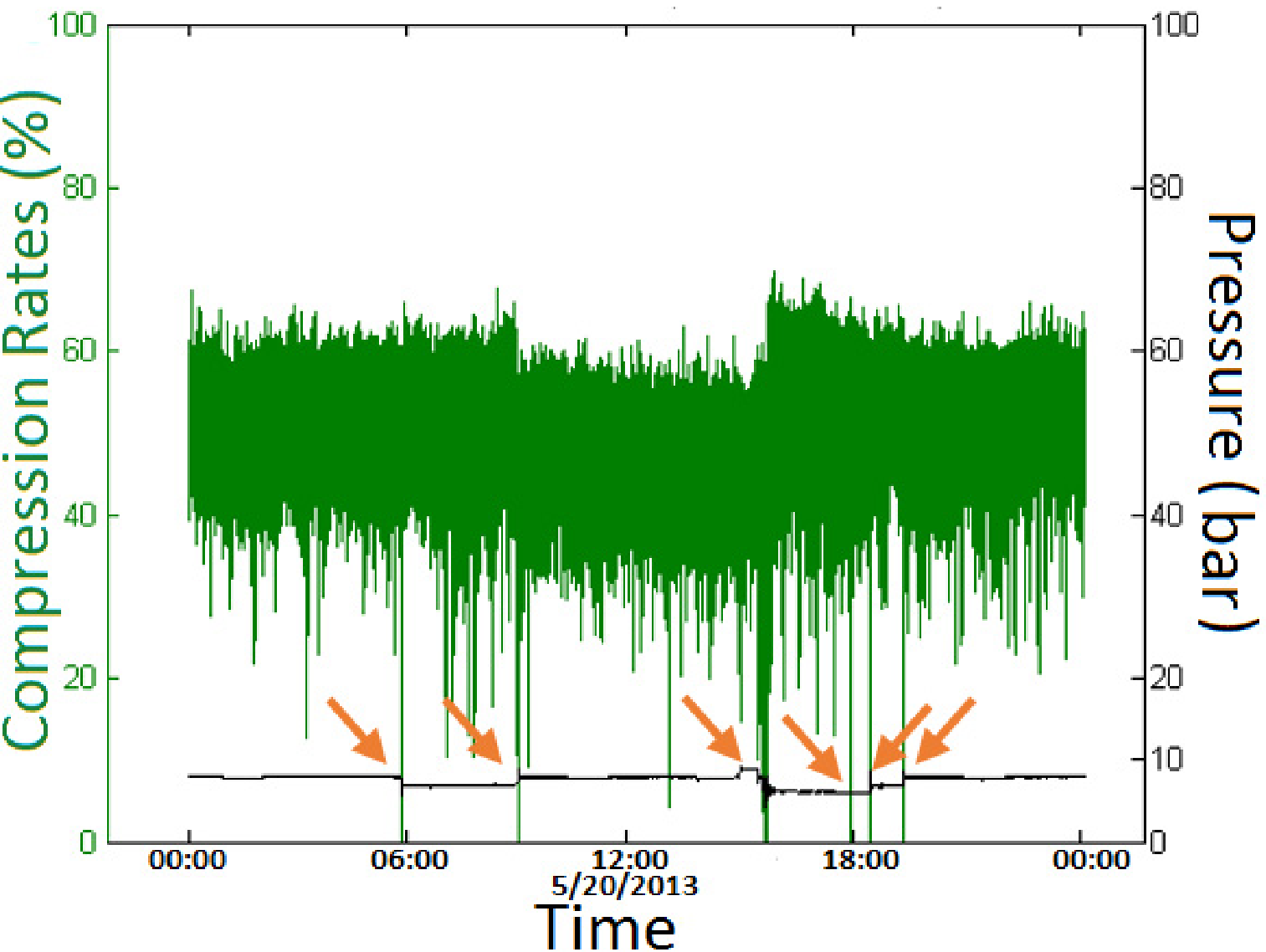}}
		\subcaption{Compression rate and raw data}
		\label{fig:Fig5a}
	\end{minipage}
	\hspace{8mm}
	\begin{minipage}[b]{0.45\textwidth}
		\centerline{\includegraphics[width=2.5in]{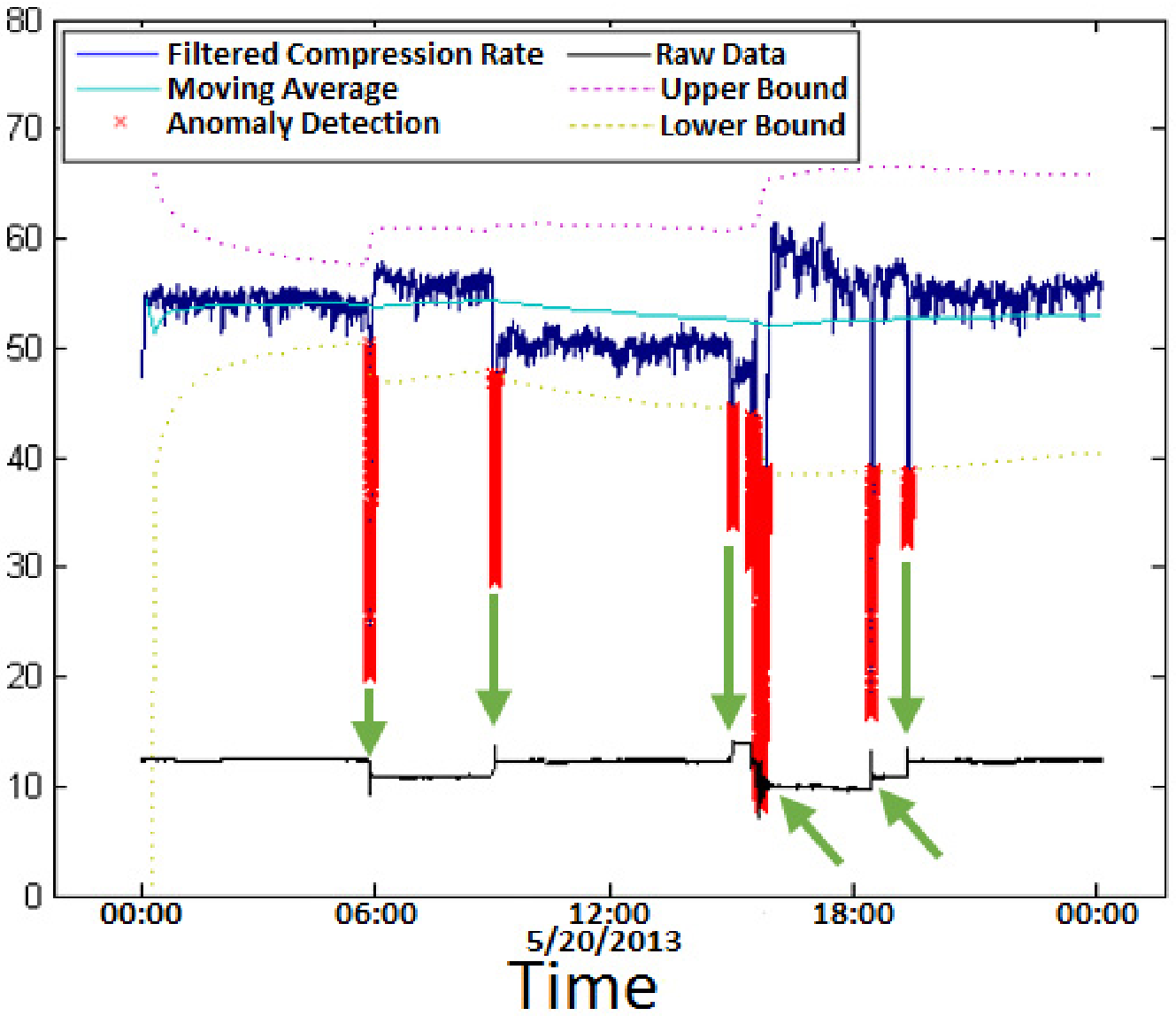}}
		\subcaption{Anomaly detection results}
		\label{fig:Fig5b}
	\end{minipage}
\caption{Observed correlation between data value fluctuation and compression rates of miniLZO algorithm, and the anomaly detection algorithm results.}
\label{fig:Fig5}
\end{figure}

\subsubsection{Raw Data Transmission Scheduling}
At each interval $t$, the set of all sensors $\mathcal{S}$ can be divided into two subsets $\mathcal{S}=\mathcal{S}_a(t)\cup\mathcal{S}_b(t)$, according whether they are forced to transmit their raw data streams or not:
\begin{itemize}
\item $\mathcal{S}_a(t)$ \textbf{(Enforce Transmission):} Each sensor $i\in \mathcal{S}_a(t)$ is enforced to transmit its raw data at $t$ in the following three cases: (1) fresh raw data from $i$ is required to update the correlation graph, (2) node $i$ has no correlation neighbor at $t$, and (3) an anomaly accouter outlier occurs at $i$. This is because when one of three cases happens, neither $i$'s own data stream nor other data streams can be reliably estimated based on the raw data of $i$. It is obvious that $y_i(t)=1,\forall i\in\mathcal{S}_a(t)$.
 \item $\mathcal{S}_b(t)$ \textbf{(Transmission Scheduling):}. Each sensor node $i\in\mathcal{S}_b(t)$ is not enforced to transmit its raw data, i.e. $y_i(t)\in\{0,1\},\forall i\in \mathcal{S}_b(t)$. Given $\mathcal{S}_b(t)$, our scheduling algorithm DTS and FAST-DTS will optimize the transmission decision, i.e. which subset of sensors in $\mathcal{S}_b(t)$ should transmit their raw data at interval $t$, based on the estimation reliability (formally defined later) and the energy states of $i$ (including battery level, harvested energy, and energy consumption). This will be discussed in detail in Section \ref{sec:Scheduling}.
\end{itemize}

Let a set of sensor nodes $\mathcal{Y}(t)\subseteq\mathcal{S}$
\begin{equation}
\mathcal{Y}(t):=\{i: i\in\mathcal{S},y_i(t)=1\}
\end{equation}
to represent the \textit{Transmission Scheduling Decision} of the whole system at interval $t$.

 \subsubsection{Data Estimation and Estimation Reliability}
The data stream of a sensor node $i$ can be estimated by its one or multiple correlation neighbors in $\mathcal{N}_i(t)$. For each sensor $i$, define its data estimation reliability as a real number between 0 and 1. Specially, if $i$ transmits its raw sensor data stream at interval $t$, its data reliability is 1. Otherwise, its data should be estimated based on the raw data stream transmitted by the other sensors correlated with it, resulting in a data reliability less than 1. Therefore, we can formalize the data reliability of each sensor node $rlb_i(\mathcal{Y}(t))$ as a function of the scheduling decision $\mathcal{Y}(t)$. The details of data stream estimation and computation of $rlb_i(\mathcal{Y}(t))$ will be presented in Section \ref{sec:estimation}. It can be seen that $rlb_i(\mathcal{Y}(t))=1, \forall i\in\mathcal{Y}(t)$, because  every sensor in set $\mathcal{Y}(t)$ is scheduled to transmit its raw data at interval $t$.  The details of data stream estimation will be discussed in Section \ref{sec:estimation}.

Table 1 summarizes the frequently used symbols in this paper.

\section{In-Node Anomaly Detection.}
\label{sec:anomaly}

In a water network, an anomaly is defined as an abrupt fluctuation in pressure data. An anomaly can be caused by: (a) a disturbance occurrence in the flow rate in the water network (e.g. leakage or burst), and (b) a hardware failure in the sensor node. In this section, we develop an algorithm to perform in-node anomaly detection  based on the data compression rate procured by a memory-efficient compression algorithm.
\vspace{0.3em}\\
\textbf{Observation 1.} \textit{A strong correlation exists between data value fluctuation and the compression rates of several compression algorithms such as S-LZW-MC \cite{sadler2006data}) and miniLZO \cite{kraus2008optimal}.}
 \vspace{0.3 em}

Intuitively, this observation holds because lossless compression algorithms reduce data volume by merging similar bytes within a packet. In the case of a stable system, the retrieved measurements fluctuate within a small range and around a certain value. This increases the similarity of the bytes in a packet, and consequently raises the compression performance. When an anomaly happens, the consecutive measurements deviate dramatically, resulting the incompressibility of the streams. Fig.~\ref{fig:Fig5} (a) shows an example of our observation. Note that the original water pressure data is also overlaid on the same graphs (lower line). It is clear from Fig.~\ref{fig:Fig5} (a), that these traces highlight data anomalies as indicated by the arrows. After multiple dataset evaluations, we formed the hypothesis that we could use the correlation of the compression rate and raw data. From water technician logs, we observed that the anomalies were valve position changes which were used to simulate water bursts, causing significant pressure data fluctuation. At these points, the compression algorithm is unable to compress the data so the compression rate falls to 0\%. In Fig.~\ref{fig:Fig5} (a), the drop in compression rate isolates the areas of raw data where the fluctuation pattern is changeable.

Based on above observation, we develop an algorithm to detect significant changes in compression rate and therefore identify the timestamps of anomalies. To achieve a high anomaly detection accuracy, noise is removed from the compression rate stream using a one-dimensional Kalman Filter \cite{kalman2009}, \cite{olfati2005distributed} indicated in Fig.~\ref{fig:Fig5}b with the upper line.

After noise removal, the anomalies can be detected accurately as shown in Fig.~\ref{fig:Fig5}b, where Kalman Filter state (upper line), raw data (lower line), and the anomalies (arrows) are illustrated. The drops are being detected by using the average $avg$ and the standard deviation $std$ of the compression rate moving average for a predefined window size $mavgw$. We use this because it smoothes the states for easier analysis and reduces threshold computation to window sizes. Specifically, every Kalman Filter value is being checked if it ranges between upper and lower bounds created by the $avg \pm std * l$, where $l$ represents the elasticity of the outlier detection (smaller values mean that the system is more sensitive. Any value that lies outside these bounds is being considered as an anomaly (Fig.~\ref{fig:Fig5}b markers).

The accuracy of the anomaly detection algorithm depends on the selection of Kalman filter parameters \cite{kalman2009}, $mavgw$, and $l$. In \cite{kartakis2014real}, we present a mechanism based on active learning notion \cite{settles2010active} for optimal parameter configuration of the proposed anomaly detection algorithm, which increases the accuracy of the detection by minimizing the False-Positives (FP) and True-Negatives. By applying the mechanism in \cite{kartakis2014real} in our evaluation data, more than 95\% accuracy was achieved.

\section{Sustainable Scheduling for Raw Data Transmission}
\label{sec:Scheduling}
In this section, we develop a theoretically optimal scheduling algorithm, DTS, which maximizes the estimation reliability while ensuring sustainable operation of the smart water sensing system. By exploiting the analytical behaviors of DTS, we then develop FAST-DTS,  a lightweight (linear complexity) and adaptive algorithm, to make the transmission scheduling decision at real time in practical smart water sensing systems.
\subsection{Monotonically Non-Decreasing Estimation Reliability}
\begin{figure}
	\centerline{\includegraphics[width=3.9in]{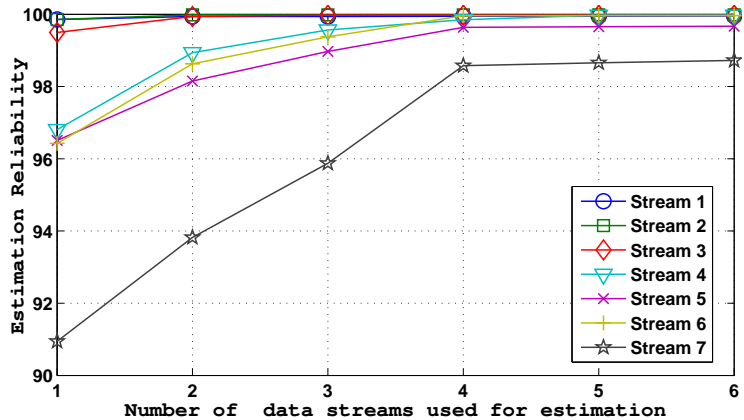}}
	\caption{results of linear regression for 1 day training set and combination size}
	\label{fig:observation1}
\end{figure}

Based on all data collected from  a real smart water sensing system, 
 we observe that the estimation reliability $rlb_i(\mathcal{Y}(t))$ has the following property:
\vspace{0.2em}\\
\textbf{Observation 2}. For any given sensor $i\in \mathcal{S}$, consider two scheduling decisions $\mathcal{Y}(t)$ and $\mathcal{Y}'(t)$, where $\mathcal{Y}'(t)=\mathcal{Y}(t)\cup\{i\}$. We have
\begin{equation}
rlb_j(\mathcal{Y}'(t))\geq rlb_j(\mathcal{Y}(t)), \forall j\in\mathcal{S}\label{eq:observation1}
\end{equation}
This observation demonstrates that adding any raw sensor data stream $i\in\mathcal{S}$ will result in a equal or larger (i.e. \textbf{a monotonically non-decreasing}) estimation reliability for all sensor data streams $j\in \mathcal{S}$. An example of this observation is illustrated in Fig.~\ref{fig:observation1}, which is based on the sensor node topology shown in Fig.~\ref{fig:Fig2}.

Inequality (\ref{eq:observation1}) is a very important property that will be used in the scheduling decision making.

\subsection{Stochastic Scheduling Optimization}
The optimal subset of transmission nodes $\mathcal{Y}(t)$ at every interval $t$ can be obtained by solving the following stochastic optimization problem:

 \begin{eqnarray}
&& \underset{\mathcal{Y}(t)}{\textbf{max}} \qquad\qquad\qquad\frac{1}{\rm t_{end}} \sum_{t=1}^{\rm t_{end}} \sum _{i\in\mathcal{S}}{U}_i(rlb_i(\mathcal{Y}(t)))
 \qquad\qquad \label{eq:opiobj}\\
&&{\rm \textbf{subject~to}}\qquad\quad \forall i\in\mathcal{S}_a(t),\forall t,~y_{i}(t)=1,rlb_i(\mathcal{Y}(t))=1 \label{eq:musttransmitting}\\
&& \qquad\qquad\qquad\qquad\forall i\in\mathcal{S}_b(t),\forall t,~y_{i}(t)\in\{0,1\},~rlb_i(\mathcal{Y}(t))\geq{\rm rlb_{min}} \label{eq:zeroonescheduling}\\
&&\qquad\qquad\qquad\qquad \forall i\in\mathcal{S},\forall t,~{\rm B_{max}}\geq B_i(t)>0 \label{eq:ENOconstraint}\\
&&\qquad\qquad\qquad\qquad \forall i\in\mathcal{S},\frac{1}{\rm t_{end}}\sum_{t=1}^{\rm t_{end}}(h_i(t)-y_{i}(t) E^{tr}_i(t)- E^{in}_i(t))\geq0\qquad\label{eq:congaurantee}
\end{eqnarray}
where the utility function $U_i(rlb_i(\mathcal{Y}(t)))$ in the objective (\ref{eq:opiobj}) can be any convex, non-decreasing, and differentiable function of $rlb_i(\mathcal{Y}(t))$. For instance, a utility function of
\begin{equation}
U_i(rlb_i(\mathcal{Y}(t)))=\ln(rlb_i(\mathcal{Y}(t))+\varepsilon)\label{eq:propofairnes}
\end{equation}
with some small value $\varepsilon>0$ can achieve approximate proportional fair reliability among sensor data streams\cite{kelly1998rate}.
Constraint (\ref{eq:musttransmitting}) highlights the sets of sensors that must transmit their archived data.
Constraint (\ref{eq:zeroonescheduling}) states the minimal correlation requirements.
Constraint (\ref{eq:ENOconstraint})  ensures that each node is not allowed to run out of battery energy or exceed its finite battery capacity.
Finally, constraint (\ref{eq:congaurantee}) demonstrates that the long-term average of consumed energy for each sensor node should not be more than that of the harvested energy.

In summary, we can see that problem (\ref{eq:opiobj})-(\ref{eq:congaurantee}) aims to maximize the long-term average estimation reliability of all sensor data streams with the consideration of fairness, while ensuring the sustainable  operation for the long-term sensing of smart water systems.
It is worth noting that fairness is an important system issue. For instance, suppose we use a simple utility function $U(rlb_i(\mathcal{Y}(t)))=rlb_i(\mathcal{Y}(t)),~\forall i,~t$, which simply maximizes the summations of all reliabilities without considering the fairness among them.
In this case, some estimated streams would have very good reliabilities (e.g. close 100\%), while others would be very poor, e.g. just reach  $\rm rlb_{min}$ in constraint in equation (\ref{eq:zeroonescheduling}). A simple way to address this issues is to introduce well-defined utility functions (e.g. the approximate proportional fair function (\ref{eq:propofairnes})), which are systematically studied in \cite{joe2013multiresource}.

\subsection {Data Transmission Scheduling (DTS)}
In this subsection, we develop an adaptive scheduling algorithm, Data Transmission Scheduling (DTS), which makes the optimal transmission decision $\mathcal{Y})(t)$ to solve the stochastic optimization problem (\ref{eq:opiobj})-(\ref{eq:congaurantee}) in real time. At the beginning of each interval $1
\leq t\leq {\rm t_{end}}$, the DTS algorithm operates as follows

1. Compute the optimal scheduling decision $\mathcal{Y}(t)$ by solving the following problem:

 \begin{eqnarray}
&&\underset{\mathcal{Y}(t)}{\textbf{max}}\qquad\quad~ \sum _{i\in\mathcal{S}}(V{U}_i(rlb_i(\mathcal{Y}(t)))+\beta_i(t)y_i(t))
 \qquad\quad \label{eq:perslotpiobj}\\
&&{\rm \textbf{subject~to}} \quad rlb_i(\mathcal{Y}(t))\geq{\rm rlb_{min}},\forall i\in\mathcal{S}\label{eq:perslotconstraint}
 \end{eqnarray}
where $V>0$ is system parameter that balances the trade-off between the system reliability performance and the risk of battery depletion, and
 \begin{eqnarray}
&&\beta_i(t)=E^{tr}_i(t)(B_i(t)-{\rm B_{exp}})
\end{eqnarray}
Here, the \textit{expected battery level} $0<{\rm B_{exp}}<{\rm B_{max}}$ is a system parameter that aims to balance the harvested energy utilization and risk of  battery depletion. We will discuss the practical settings of $V$ and ${\rm B_{exp}}$ in detail in the next subsection.

2. Update the battery of each sensor node $i$ using Eq.(\ref{eq:energybufferupdate}).
If a battery overflow occurs, i.e. $B_i(t+1)>{\rm B_{max}}$, then set $B_i(t+1)={\rm B_{max}}$.

It is worth noting that the DTS algorithm \textit{only requires} current system knowledge, when can be obtained easily at real time, and \textit{does not} need to predict the information of future intervals, such as future energy harvesting opportunities and data stream correlations.

\subsection{Analytical Results and Parameter Settings}
The analytical results summarized by Theorem 1 below will demonstrate that the DTS algorithm has strong performance guarantees, and provides guiding principles for setting system parameters $V$ and ${\rm B_{exp}}$ in practice.
\vspace{1em}
\\
\textbf{Theorem 1.} \textit{DTS algorithm has the following performance guarantees.}
\begin{enumerate}
\item\textbf{Sustainable Operation.} By setting
\begin{equation}
V\leq V_{threshold}=\frac{\sum _{i\in\mathcal{S}}(U_i(1)-U_i(0))}{\rm E_{max}B_{exp}}\label{ieq:Vthreshold}
\end{equation}
DTS will make a real-time scheduling decision $\mathcal{Y}(t)$ such that every node $i$ will not transmit its raw data stream when its battery level $B_i(t)<{\rm B_{exp}}$, unless its reliability constraint is not satisfied, i.e.
$rlb_i(\mathcal{Y}(t))<{\rm rlb_{min}}$. Here $\rm E_{max}$ is the upper bound of energy costs for transmission for all intervals (e.g. $\forall t,i, E^{tr}(t)\leq{\rm E_{max}}$), which is a constant value depending on the wireless transceiver adopted by the sensing device.
\item\textbf{Asymptotic Optimality}. \textit{As $V\rightarrow +\infty$, DTS  asymptotically achieves the optimal solution to the stochastic optimization problem  (\ref{eq:opiobj})-(\ref{eq:congaurantee}).}
\end{enumerate}
\textbf{Proof.}
Theorem 1 can be proved by using sample-path based Lyapunov optimization theory, which can be found in the Appendix.\hfill$\Box$

We first discuss the impact of $V$ on the behaviors of DTS algorithm. With a fixed ${\rm B_{exp}}$, a larger $V$ will result in better reliability performance, but  more aggressive energy consumption behavior. Especially, when $V$ is larger than the threshold $V_{threshold}$ defined by the right-hand side of inequality (\ref{ieq:Vthreshold}), a sensing device would be forced to transmit its raw data, even when it has a low battery energy level i.e. smaller than the ${\rm B_{exp}}$.   In order to achieve sustainable operation, we can set $V= V_{threshold}$ in practice.

The expected battery level ${\rm B_{exp}}$ also has a significant impact on the DTS performance. By setting a larger ${\rm B_{exp}}$, the risk of battery depletion will be lower, but the probability of battery overflow will be higher, resulting in lower harvested energy utilization. In addition, larger ${\rm B_{exp}}$ will lead to a small $V_{threshold}$, and therefore, lower reliability performance. In practice, we can set ${\rm B_{exp}}$ according to the long-term energy harvesting and consumption behaviors of the smart water system. For instance, if the correlation graph is always dense, overtime (good estimation opportunity and low energy consumption), we can use a small ${\rm B_{exp}}$ to improve the system performance. However, if the correlation graph is sparse or evenly disconnected within some durations (i.e. poor estimation opportunity and large energy consumption to ensure constraint (\ref{eq:perslotconstraint}) within such durations), we need to set a large ${\rm B_{exp}}$ to reduce battery depletion risk within such durations.

\begin{figure}
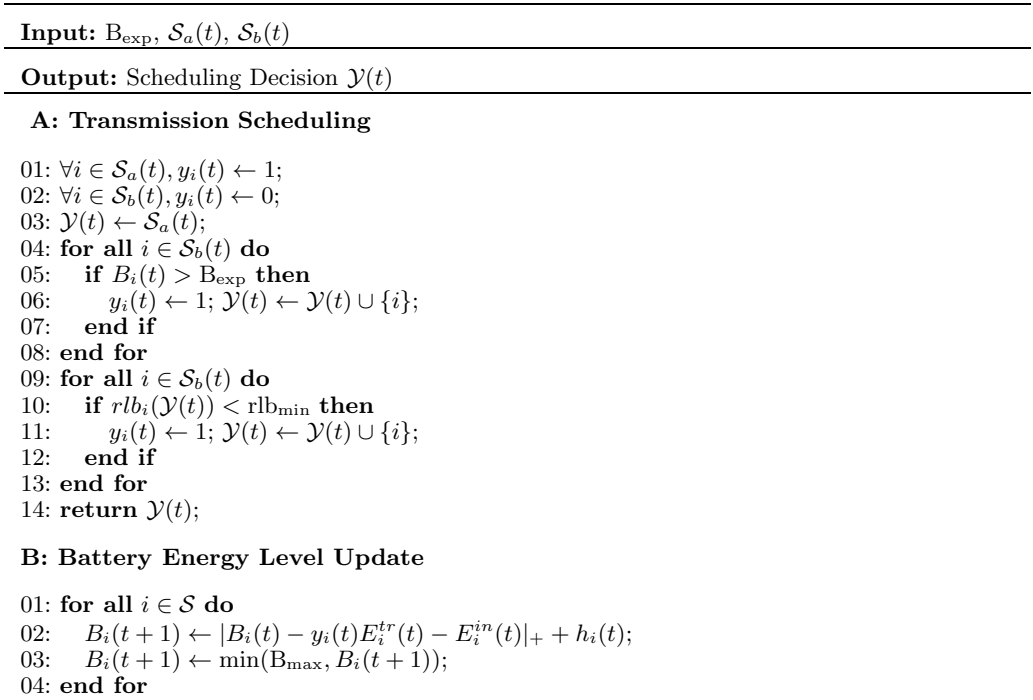

 \fontsize{9 pt}{\baselineskip}\selectfont
  \renewcommand{\arraystretch}{0.9}
  \begin{tabular}[htbp]{p{5.2in}}
  \hline
  \vspace{0.001em}
\textbf{Input:} $ {\rm B_{exp}}$, $ \mathcal{S}_a(t)$, $ \mathcal{S}_b(t)$\\
  \hline
  \vspace{0.001em}
\textbf{Output:} Scheduling Decision $\mathcal{Y}(t)$\\
  \hline
  \vspace{0.001em}
\textbf{ A:  Transmission Scheduling} \\
  \vspace{0.2em}
01:~$\forall i\in \mathcal{S}_a(t), y_i(t)\leftarrow1$; \\
02:~$\forall i\in \mathcal{S}_b(t),y_i(t)\leftarrow 0$;\\
03:~$\mathcal{Y}(t)\leftarrow \mathcal{S}_a(t)$; \\
04:~\textbf{for all} $i\in \mathcal{S}_b(t)$ \textbf{do}\\
05:~\quad\textbf{if} $ B_i(t)>{\rm B_{exp}}$ \textbf{then}\\
06:~\quad\quad $y_i(t)\leftarrow1$;~$\mathcal{Y}(t)\leftarrow\mathcal{Y}(t)\cup\{i\}$;\\
07:~\quad\textbf{end if} \\
08:~\textbf{end for} \\
09:~\textbf{for all} $i\in \mathcal{S}_b(t)$ \textbf{do}\\
10:~\quad\textbf{if} $ rlb_i(\mathcal{Y}(t))<{\rm rlb_{min}}$ \textbf{then}\\
11:~\quad\quad $y_i(t)\leftarrow1$;~$\mathcal{Y}(t)\leftarrow\mathcal{Y}(t)\cup\{i\}$;\\
12:~\quad\textbf{end if} \\
13:~\textbf{end for} \\
14:~\textbf{return} $\mathcal{Y}(t)$;\\
  \vspace{0.2em}
\textbf{B:  Battery Energy Level Update} \\
  \vspace{0.2em}
01:~\textbf{for all} $i\in\mathcal{S}$ \textbf{do}\\
02:~\quad$B_i(t+1)\leftarrow|B_i(t)-y_{i}(t)E^{tr}_i(t)- E^{in}_i(t)|_{+}+h_i(t)$;  \\
03:~\quad$B_i(t+1)\leftarrow \min({\rm B_{max}}, B_i(t+1) )$;\\
04:~\textbf{end for}\\
\hline
\end{tabular}
\caption{Pseudocode of FAST-DTS at every interval $t$.}
\label{fig:FAST-DTS}
\end{figure}

\subsection {NP Completeness of the DTS}

At each interval $t$, DTS needs to compute the optimal $\mathcal{Y}(t)$ to solve the deterministic optimization problem (\ref{eq:perslotpiobj})-(\ref{eq:perslotconstraint}), which looks very simple but may introduce extremely intensive computation. In the worst case, when $\mathcal{S}_b(t)=\mathcal{S}$ and the correlation graph $G(\mathcal{S},\mathcal{L}(t))$ is a complete graph (i.e. All sensor data streams are strongly correlated with each other), the number of all possible scheduling decisions is $2^{|\mathcal{S}|}-1$. This is because that $\mathcal{Y}(t)\subseteq\mathcal{S}$, and all possible non-empty subsets of $\mathcal{S}$ should be considered as the best one that solves problem (\ref{eq:perslotpiobj})-(\ref{eq:perslotconstraint}).
\vspace{0.2em}\\
Theoretically, we have
\\\textbf{Theorem 2.} \textit{Problem (\ref{eq:perslotpiobj})-(\ref{eq:perslotconstraint}) is NP complete}.
\vspace{0.2em}\\
\textbf{Proof.}
Since the variable $y_i(t)$ is a binary variable for each sensor node $i$, i.e. $y_i(t)\in\{0,1\}, \forall i$,  Problem (\ref{eq:perslotpiobj})-(\ref{eq:perslotconstraint}) is a mixed 0-1 programming problem, which is well-known to be NP complete in general~\cite{cormen2001introduction}.\hfill$\Box$

As a result, although the DTS algorithm can guarantee an asymptotically optimal solution to problem (\ref{eq:opiobj})-(\ref{eq:congaurantee}), computing the optimal scheduling policy at each interval is prohibitive for large-scale smart water sensing systems. To address this issue, we develop FASR-DTS, an approximate algorithm of DTS.

\subsection{FAST-DTS A Linear Approximation of DTS}

FAST-DTS has a linear complexity of $O(|\mathcal{S}|)$, based on the discussions of practical parameter settings presented in the last subsection. The pseudo code of FAST-DTS is shown in Fig. \ref{fig:FAST-DTS}.

From Fig.\ref{fig:FAST-DTS}, we can see that FAST-DTS implicitly sets the parameter $V=V_{threshold}$: If the battery level of a node is higher than $\rm B_{exp}$, it will be scheduled to transmit its data stream at current interval (lines 04-08, part A); otherwise, it will not transmit unless its reliability constraint is not satisfied (lines 09-13, part A).

The operations of DTS and FAST-DTS are different in the case were the reliability constraints for the nodes with battery levels lower than $\rm B_{exp}$ are not satisfied. In this case, DTS will maximize objective (\ref{eq:perslotpiobj}) by considering all possible scheduling policies (which introduces exponential worst-case complexity), while FAST-DTS simply forces these nodes to transmit one by one until all of their reliability constraints are satisfied, according to the observed property (\ref{eq:observation1}). Although this aggressive energy consumption behavior of FAST-DTS would result in slightly higher risk of battery depletion, the system reliability performance will be better. More importantly, the computational complexity would be dramatically reduced, which enables the real-time application of FAST-DTS for large-scale smart water sensing systems.

\section{Data Stream Estimation}
\label{sec:estimation}

In this section, we discuss how to estimate the data stream of a given sensor node $i$ based on the a given set of sensors $\mathcal{Y}(t)\subseteq\mathcal{S}$, and compute the estimation reliability $rlb_i(\mathcal{Y}(t))$.

Our estimation is based on multiple auto-regressive models\cite{papadimitriou2013dimensionality}, such as linear, quadratic, pure-quadratic, and polynomial model.
For simplicity, we use linear multiple regression models to discuss our estimation approach. Let two vectors $\textbf{x}_i(t)$ and $\widehat{\textbf{x}}_i(t)$ be the raw and estimated data streams of sensor $i$ at interval $t$ respectively. The fitted multiple linear regression models have the form:
\begin{equation}
\hat{\textbf{x}}_i(t) = b_0(w) + b_1(w)\textbf{x}_{1}(t) + ... + b_n(w)\textbf{x}_{n}(t)
\end{equation}
where $n=|\mathcal{Y}\cap\mathcal{N}_i(t)|$ is the number of raw data streams that are used for the estimation, because only the correlated neighbors of sensor $i$ can be used to estimate its data;
 $b_0(w),...b_{n}(w)$ are the regression coefficients trained based on the received data streams with a window size of $w$; $\textbf{x}_1(t),...,\textbf{x}_n$ are raw data streams of $i$'s correlation neighbors in $\mathcal{Y}\cap\mathcal{N}_i(t)$.

Let a n+1 dimensional vector $\textbf{b}(w)=(b_0(w),...b_{n}(w))$ to represent all the regression coefficients. The Sum of Squares of due to Error (SSE)
 \begin{equation}
SSE (\textbf{b}(w)) =(\|\textbf{x}_i(t)-\hat{\textbf{x}}_i(t)\|_2)^2
\end{equation}
is a function of $\textbf{b}(w)$, where $\|.\|$ is the 2-norm operator. Therefore, it is easy to compute the optimal regression coefficients
 \begin{equation}
\textbf{b}^{opt}(w)=\arg\min_{\textbf{b}(w)} SSE (\textbf{b}(w))
\end{equation}

We use the coefficient of determination to measure the reliability of the estimated data stream $\hat{\textbf{x}}_i(t)$ based on $\mathcal{Y}$:
\begin{equation}
rlb_i(\mathcal{Y}(t))=1-\frac{SSE}{SST}=1-\frac{(\|\textbf{x}_i(t)-\hat{\textbf{x}}_i(t)\|_2)^2}{(\|\textbf{x}_i(t)-\overline{\textbf{x}}_i(t)\|_2)^2}\label{eq:Rsquare}
\end{equation}
 Where the total corrected sum of squares $SST =(\|\textbf{x}_i(t)-\overline{\textbf{x}}_i(t)\|_2)^2$ and $\overline{\textbf{x}}_i(t)$ is the mean of all values in vector $\textbf{x}_i(t)$.

\section{Evaluation}
\label{sec:evaluation}

In this section, we evaluated the performance of our FAST-DTS algorithm by using 170-day water pressure (128 samples per second) and energy harvesting data from our real smart water system, which consists of 24 sensor nodes. Additionally, we verified the adaptive behaviour of the system in anomalies by using and extending a small-scale testbed, WaterBox \cite{kartakis2015waterbox}. The hardware infrastructure of WaterBox is based on Intel Edison which retrieves high sample rate pressure, flow, temperature, and energy consumption data. In spite of the capability of battery connection, the sensor nodes are directly connected to power and the battery is virtualized to ensure the stability of experiments.

\subsection{Evaluation Set-up}
Since no prior related algorithm exists, we compared the performance of our FAST-DTS (FDTS) with the following three naive scheduling algorithms:
\begin{itemize}
\item \textbf{Reliability-Greedy(RG):} All the sensor nodes transmit data at every interval to achieve the highest reliability. Aforementioned state of the art solutions \cite{allen2013water,i2o2015,derceto2015} exploits this algorithm.
\item \textbf{Energy-Greedy (EG\textit{m}):} The $m$ sensor nodes with the highest battery levels transmit raw data.
\item \textbf{Round Robin (RR\textit{m}):} The selection of sensor nodes is shifted by $m$ nodes at every interval.
\end{itemize}

In all evaluations, we set the duration of an interval and training window size as 15 (commonly used by utility companies) and 60 minutes (heuristically found) respectively. In order to perform a strict benchmark in terms of energy consumption, we selected to use smaller batteries with $\rm B_{max} = 500J$ instead of the real battery used in our smart water system (i.e. 60.2 kJ). Based on WaterBox sensing device hardware \cite{kartakis2015waterbox} and the real experiments of LPWA communication technologies as conducted in \cite{kartakis2016demystifying}, the energy consumption for in-node operations $E^{in}_i(t)$ and $E^{tr}_i(t)$ for each node $i$ at every interval $t$ are set as random variables with averages of 6 J and 30 J respectively.

\begin{figure}
\centering
	\begin{minipage}[b]{0.45\textwidth}
		\centerline{\includegraphics[width=2.4in]{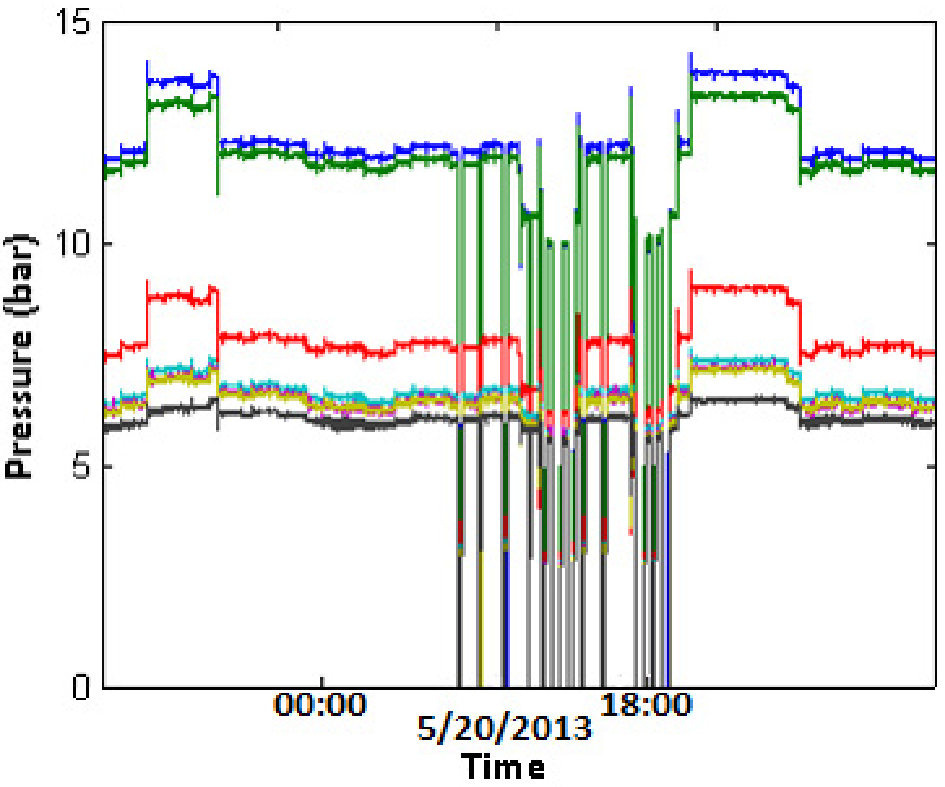}}
		\subcaption{Pressure data transmission using RG}
	\end{minipage}
	\hspace{4mm}
	\begin{minipage}[b]{0.45\textwidth}
		\centerline{\includegraphics[width=2.7in]{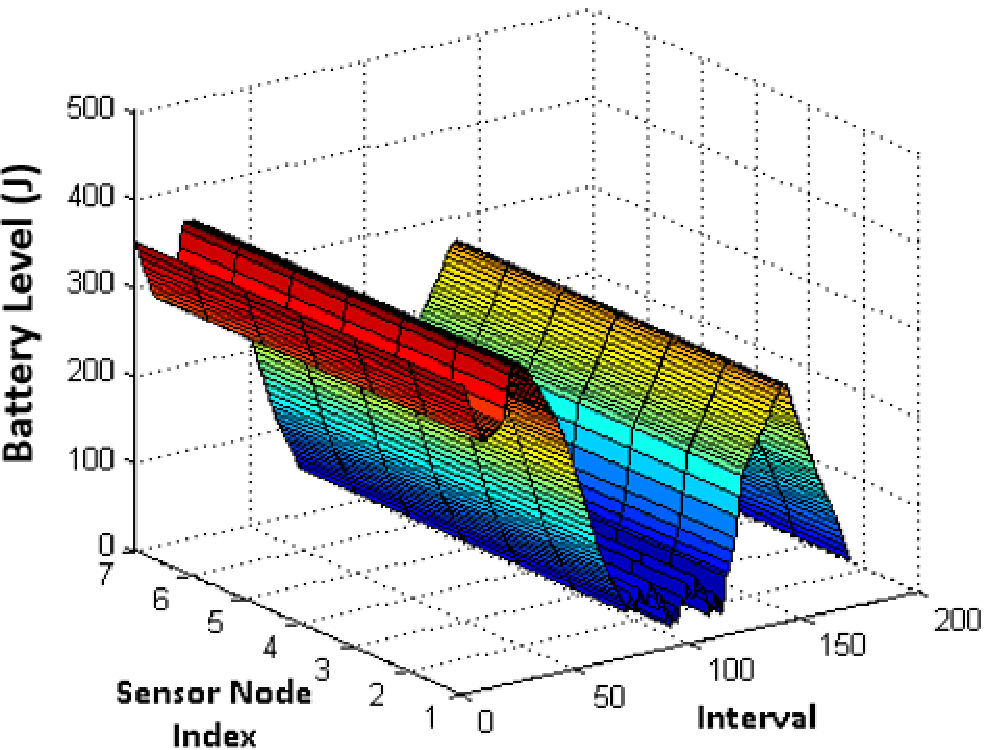}}
		\subcaption{Battery level evolution using RG}
	\end{minipage}
	\begin{minipage}[b]{0.45\textwidth}
		\centerline{\includegraphics[width=2.4in]{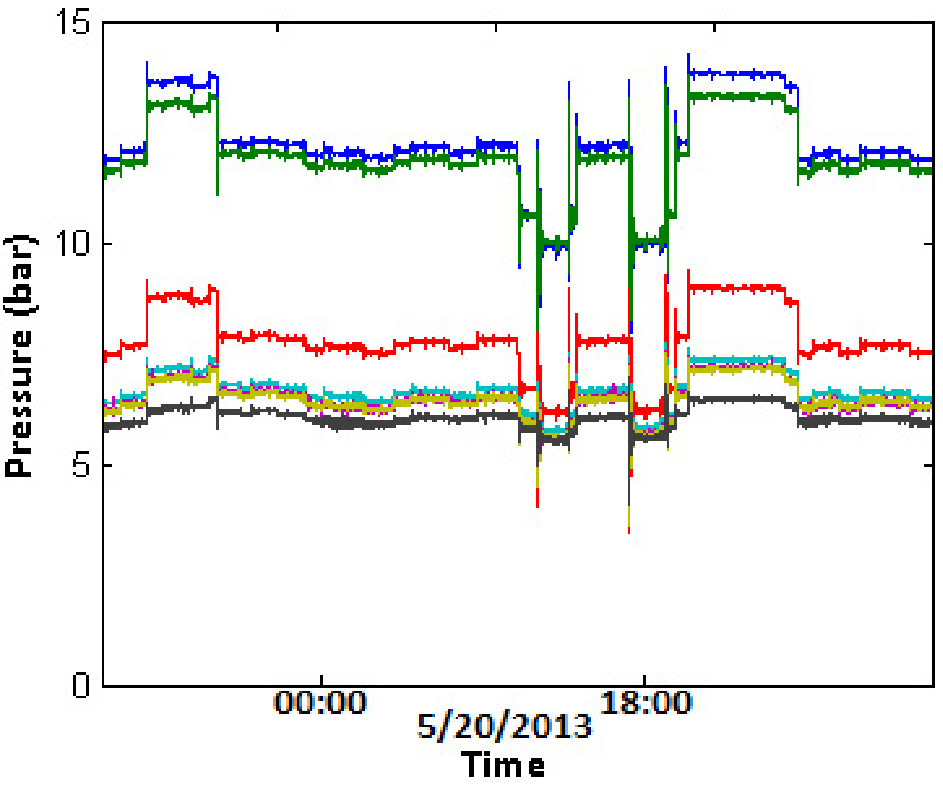}}
		\subcaption{Pressure data transmission using FAST-DTS}
	\end{minipage}
	\hspace{4mm}
	\begin{minipage}[b]{0.45\textwidth}
		\centerline{\includegraphics[width=2.7in]{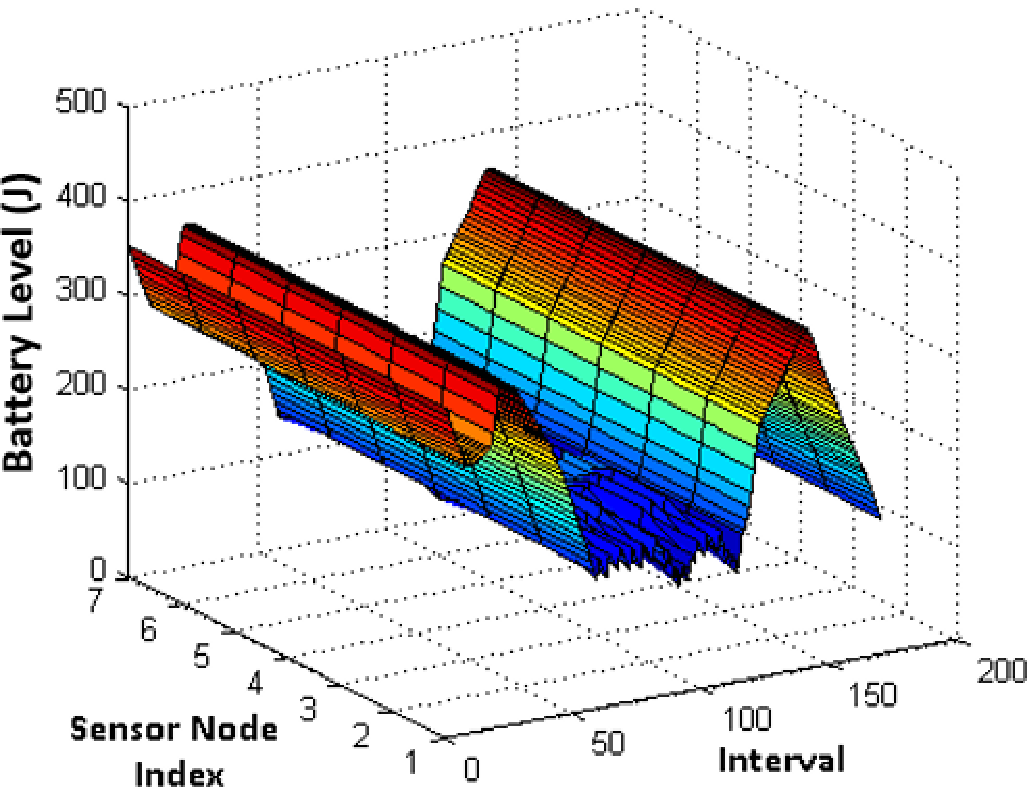}}
		\subcaption{Battery level evolution using FAST-DTS}
	\end{minipage}
	\caption{Reliability-Greedy and FAST-DTS comparison based on reliability and battery level per node}
	\label{fig:Fig11}
\end{figure}

\subsection{A case study}

In this subsection, we present a simple case study of 7 highly-correlated sensor data streams with fixed system parameter settings ($\rm B_{exp}=150$ and $\rm rlb_{min} = 0.98$) to illustrate the behaviour of the above algorithms. For brevity, the following results refer to one set of highly correlated sensor nodes\footnote{The largest correlated set of our infrastructure consists of seven nodes, which is selected to represent our simulation results. In addition, four more sets of correlated nodes were considered, which consisted of 3, 3, 5, and 6 sensor nodes respectively.}, while $m=3, 6$ for Energy-Greedy (EG3, EG6) and Round Robin (RR3, RR6).

To examine data reliability results, we introduce a new metric which was applied only to test datasets for every interval and not during training. Similar to Eq.(\ref{eq:Rsquare}) in Section \ref{sec:estimation}, instead of using SSE, we calculate $(\|\textbf{x}_i(t)-\hat{\textbf{x}}_{i,-i}(t)\|_2)^2$, where $\hat{\textbf{x}}_{i,-i}(t)$ are the estimation of $\textbf{x}_i(t)$ in the test set.

\begin{table}
\centering
\caption{Evaluation results of the case study.}
\begin{tabular}{lcccl}
\hline
\multicolumn{1}{|c|}{Algorithm} & \multicolumn{1}{c|}{\begin{tabular}[c]{@{}c@{}}Reliability\\ (\%)\end{tabular}} & \multicolumn{1}{c|}{\begin{tabular}[c]{@{}c@{}}Energy\\ Waste (kJ)\end{tabular}} & \multicolumn{1}{c|}{\begin{tabular}[c]{@{}c@{}}Transmission\\Gaps\end{tabular}}\\ \hline
\multicolumn{1}{|l|}{RG} & \multicolumn{1}{c|}{92.8} & \multicolumn{1}{c|}{0.5} & \multicolumn{1}{c|}{84}\\ \hline
\multicolumn{1}{|l|}{EG3} & \multicolumn{1}{c|}{31.9} & \multicolumn{1}{c|}{15.3} & \multicolumn{1}{c|}{0}\\ \hline
\multicolumn{1}{|l|}{EG6} & \multicolumn{1}{c|}{99.1} & \multicolumn{1}{c|}{1.4} & \multicolumn{1}{c|}{0}\\ \hline
\multicolumn{1}{|l|}{RR3} & \multicolumn{1}{c|}{69.3} & \multicolumn{1}{c|}{15.4} & \multicolumn{1}{c|}{0}\\ \hline
\multicolumn{1}{|l|}{RR6} & \multicolumn{1}{c|}{99.3} & \multicolumn{1}{c|}{1.4} & \multicolumn{1}{c|}{0}\\ \hline
\multicolumn{1}{|l|}{FDTS} & \multicolumn{1}{c|}{99.8} & \multicolumn{1}{c|}{0.5} & \multicolumn{1}{c|}{0}\\ \hline
\end{tabular}
\label{table:TableRes}
\end{table}

To provide an easier interpretation of the reliability and to define the notion and relationship with transmission gaps, we present detailed results from one and a half day. Furthermore, the selected time window is highly representative due to the pattern repeatability of each pressure stream within a water network (see Fig.\ref{fig:Fig11}). As shown in Table \ref{table:TableRes} FAST-DTS achieves the highest estimation reliability over all algorithms. Furthermore, although the expected reliability of estimation models is above 95\% for all nodes, the actual estimation results of some algorithms are less. This is caused by two reasons: (1) the non-optimal selection of a small set of nodes to transmit raw data, and (2) the transmission gaps. For example, EG3 and RR3 belong to the first case and produce the lowest estimation reliability results, 31.9\% and 69.3\% respectively. For the second case, the RG algorithm performs 7\% less in terms of estimation reliability, in spite of the continuous transmission from all the nodes. The reason is that in some intervals the nodes are unable to transmit data, because of battery depletion and transmission gaps shown in last column of Table \ref{table:TableRes}. These gaps equate data reliability to 0\% for these intervals (since no data can be transmitted). Fig.\ref{fig:Fig11}a and Fig.\ref{fig:Fig11}¬c illustrate the transmitted data from RG and FAST-DTS respectively and emphasize the transmission gaps of RG (zero values in Fig.\ref{fig:Fig11}¬a).

During transmission gaps, RG aggressively utilize energy and forced each node's battery level to zero (Fig.\ref{fig:Fig11}¬b), leading to unsustainability of the system and frequent transmission gaps (Fig.\ref{fig:Fig11}¬a). In contrast, FAST-DTS trys to keep the battery level above $\rm B_{exp}$, as shown in Fig.\ref{fig:Fig11}¬ d respectively. This successfully avoids transmission gaps and therefore system sustainability. In addition, FAST-DTS also achieves the best performance in harvested energy utilization. In some cases, the energy harvesting system can produce more energy than the sensor nodes' need and the system wastes energy. For example, the EG3 and RR3 algorithms which select only three nodes per interval keeps the energy close to $\rm B_{max}$ and at every interval waste energy that could be used to schedule more nodes to transmit raw data. In summary,  FAST-DTS performs the best battery management by minimizing energy waste and avoiding transmission gap, as shown in  Table¬\ref{table:TableRes}.

\subsection{System Adaptation in Anomalies}

\begin{figure}
	\centerline{\includegraphics[width=3.5in]{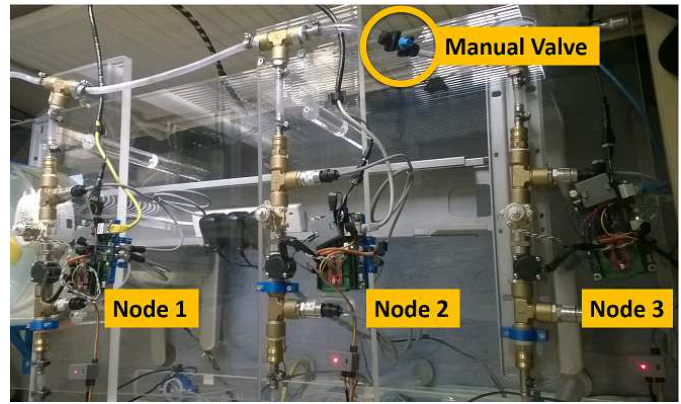}}
	\caption{Testbed modification for anomaly emulation.}
	\label{fig:Figwm}
\end{figure}

\begin{figure}
	\centerline{\includegraphics[width=3.5in]{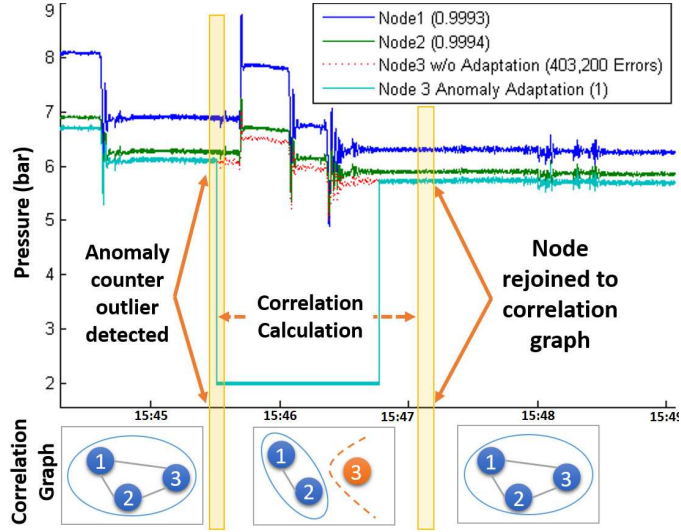}}
	\caption{Correlation adaptation based on anomaly variance.}
	\label{fig:Fig12}
\end{figure}

In this subsection, we present the impact of an observed anomaly on the behaviour of our system and the correlation graph update. As shown in Fig.~\ref{fig:Figwm}, we modify WaterBox to emulate the anomaly behaviors of  three DMAs' sensor nodes~(Nodes 1-3). Specifically, a manual valve was installed between the second and third DMAs, and bursts were emulated by opening/closing the valve abruptly.

Fig.\ref{fig:Fig12} presents an example of these three sensor nodes and the system's behavior in an anomaly which was occurred due to abrupt closing of the valve before Node 3. During the closing of the valve (Fig.\ref{fig:Fig12} first vertical line), the three sensor nodes transmit to the data processing center anomaly counters 3, 2, and 100 respectively due to sudden pressure drop in Node 3. Then, the data center detects the Node 3 anomaly counter as an outlier and forces Node 3 to transmit raw data. After receiving raw data, the data processing center calculates the correlation and decides to remove Node 3 from the correlation graph. Without this functionality, Node 3's data could continue to be estimated by other nodes' data and to produce inaccurate data (horizontal dashed line arrows - 403200 wrong estimation values to the data center). After some intervals (second vertical line), we open the valve and Node 3 returns to normal hardware operation. Then, the data processing center observes high correlation and rejoins Node 3 to the correlation graph.

\subsection{The Impacts of System Parameter Settings}
\begin{figure}
	\centering
	\begin{minipage}[b]{0.32\textwidth}
		\centerline{\includegraphics[width=\textwidth]{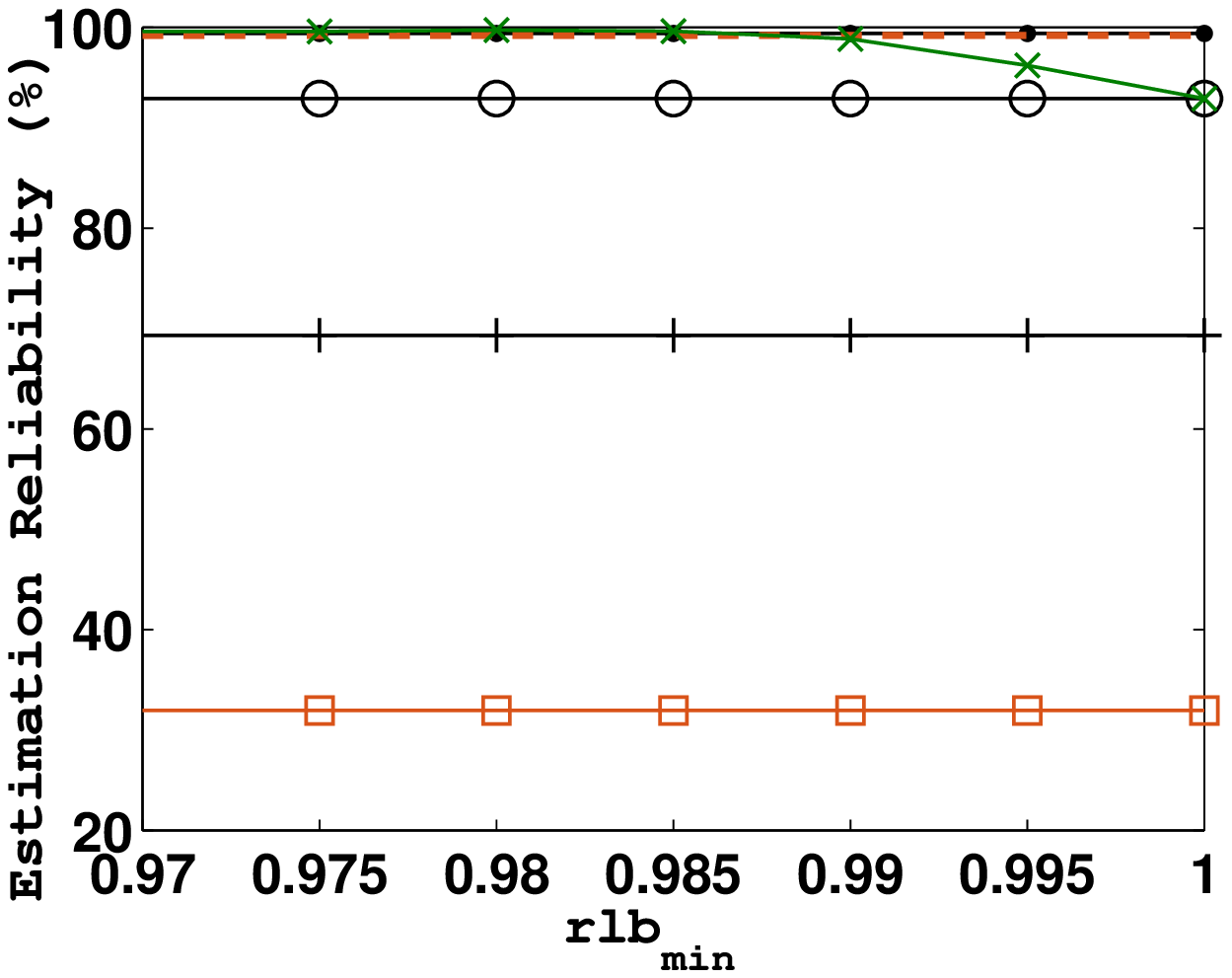}}
		\subcaption{estimation reliability}
	\end{minipage}
	\begin{minipage}[b]{0.32\textwidth}
		\centerline{\includegraphics[width=\textwidth]{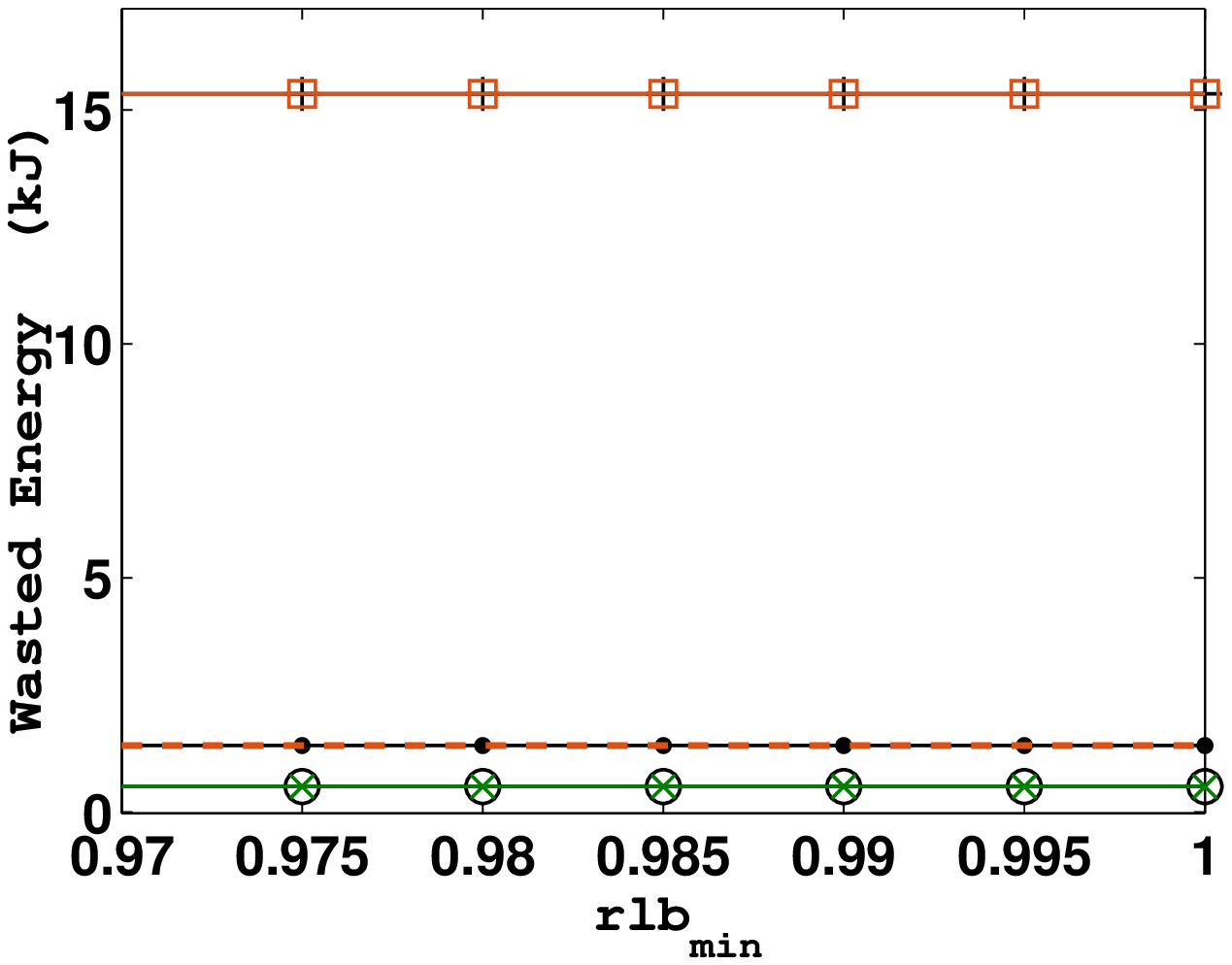}}
		\subcaption{wasted energy}
	\end{minipage}
	\begin{minipage}[b]{0.32\textwidth}
		\centerline{\includegraphics[width=\textwidth]{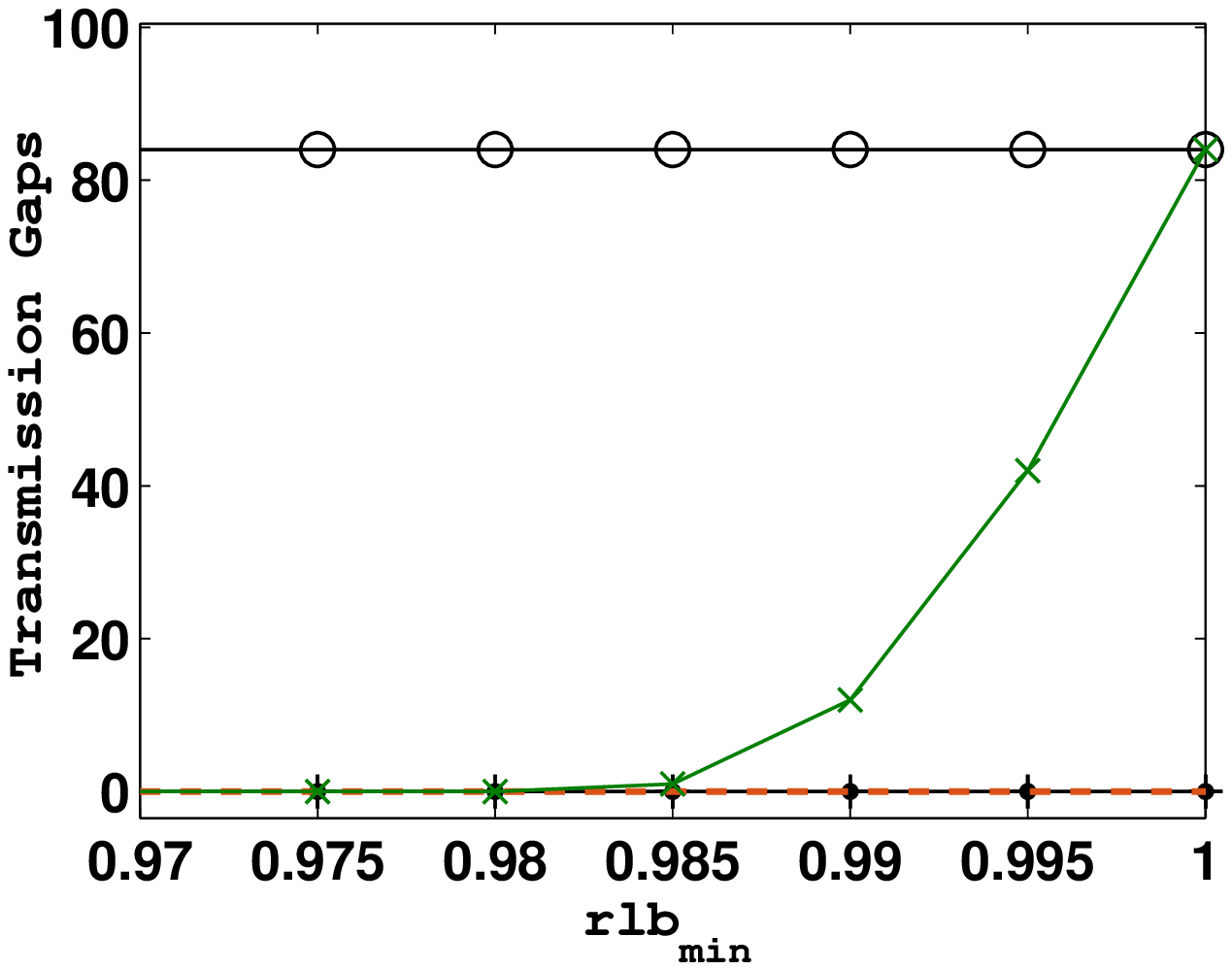}}
		\subcaption{transmission gaps}
	\end{minipage}

	\begin{minipage}[b]{0.32\textwidth}
		\centerline{\includegraphics[width=\textwidth]{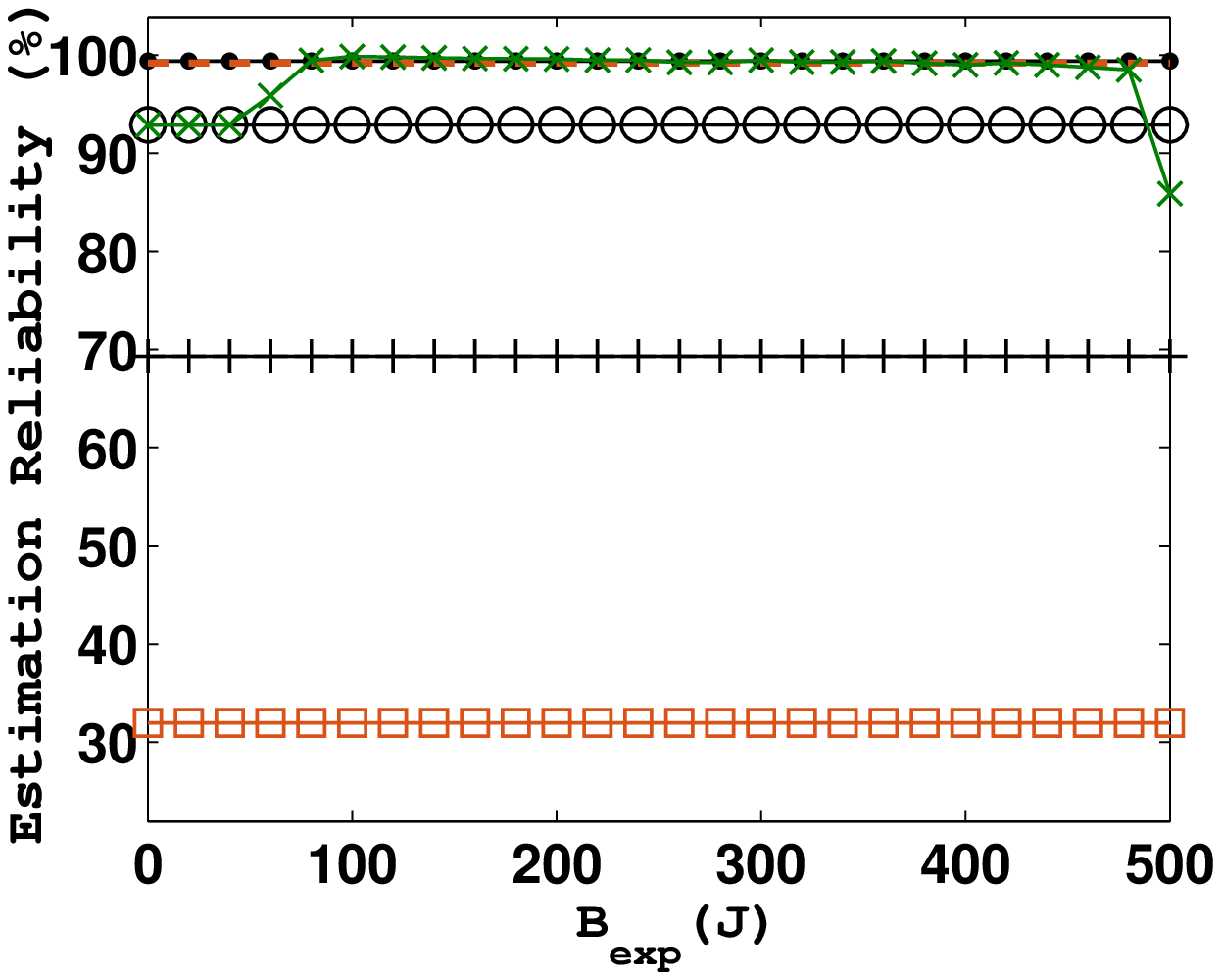}}
		\subcaption{estimation reliability}
	\end{minipage}
	\begin{minipage}[b]{0.32\textwidth}
		\centerline{\includegraphics[width=\textwidth]{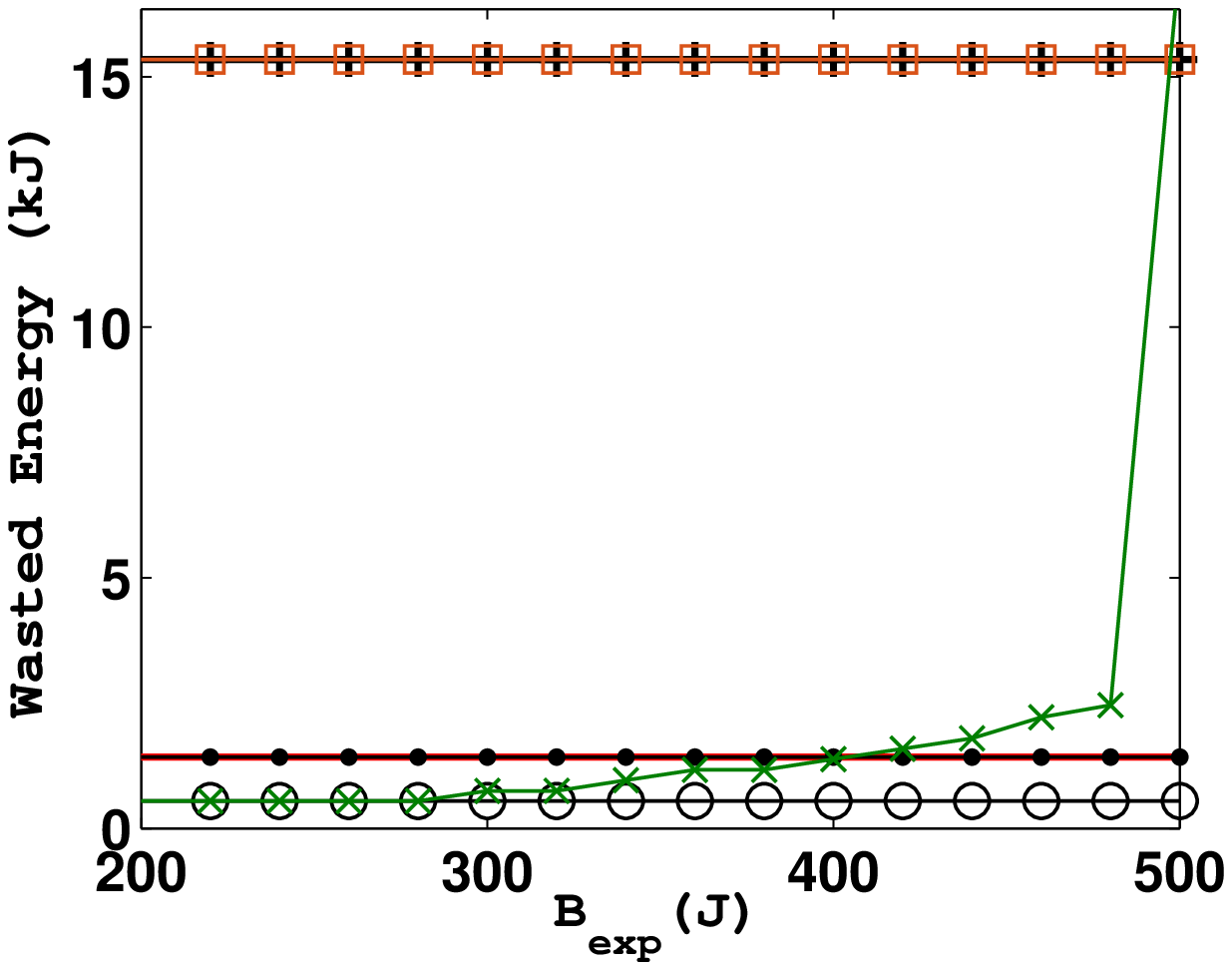}}
		\subcaption{wasted energy}
	\end{minipage}
	\begin{minipage}[b]{0.32\textwidth}
		\centerline{\includegraphics[width=\textwidth]{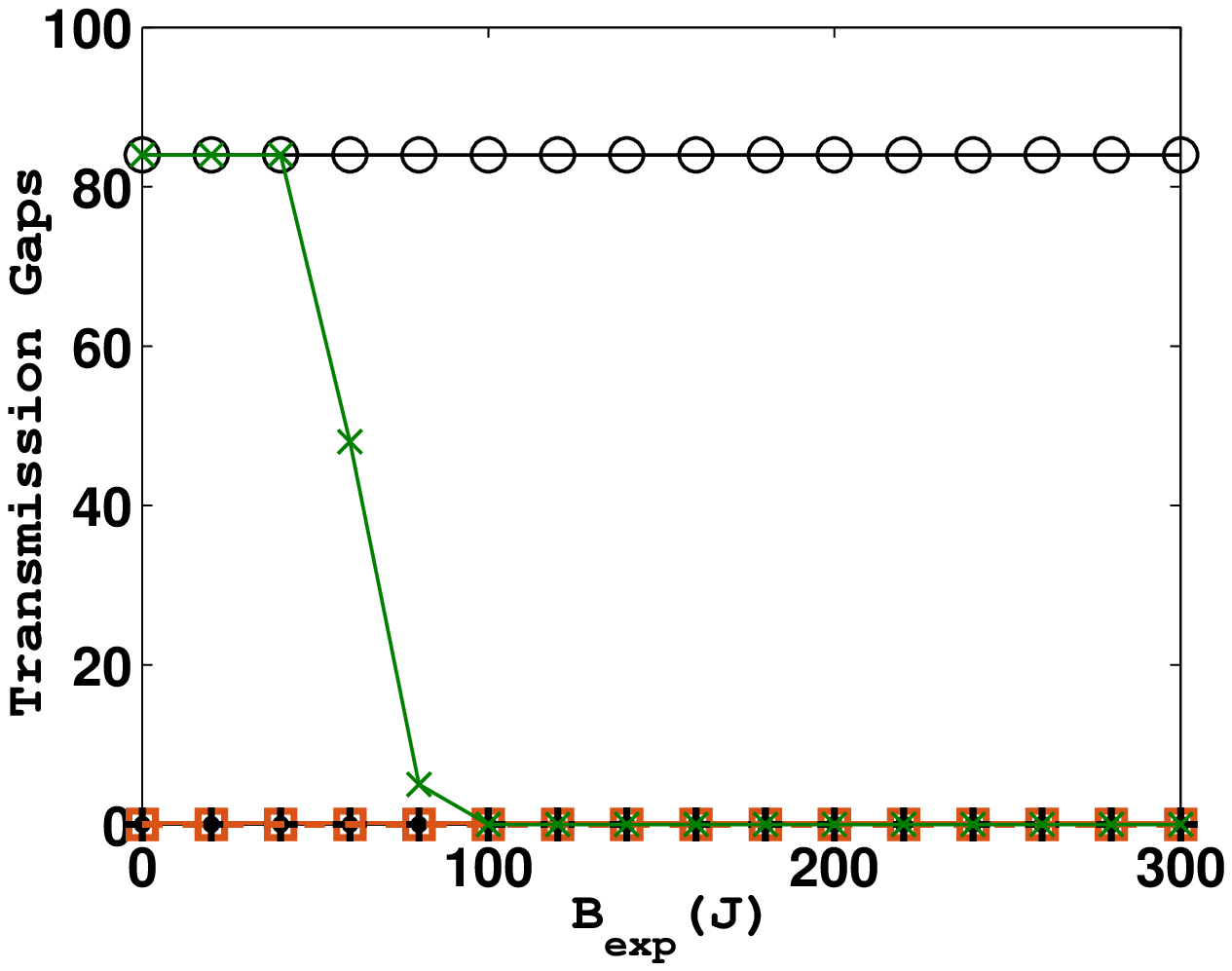}}
		\subcaption{transmission gaps}
	\end{minipage}
	
	\begin{minipage}[b]{0.45\textwidth}
		\centerline{\includegraphics[width=\textwidth]{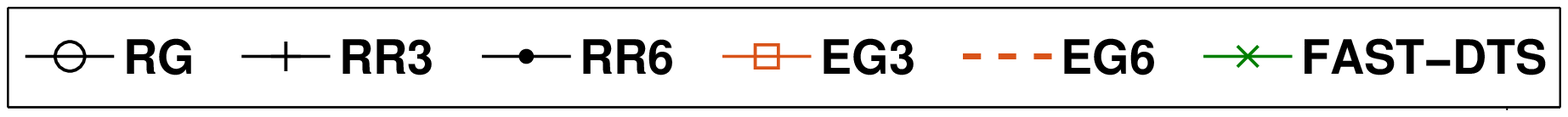}}
	\end{minipage}
	\caption{Reliability, energy waste, and transmission gaps with different $\rm rlb_{min}$ and $\rm B_{exp}$.}
	\label{fig:Fig11r}
\end{figure}

The system parameters $\rm B_{exp}$ and $\rm rlb_{min}$ have significant impacts on our smart water system, as shown in Fig.\ref{fig:Fig11r}. As the $\rm rlb_{min}$ tends to 1, the FAST-DTS behaves like RG. This is because that more nodes are forced to transmit raw data to achieve a 100\% reliability, resulting in the increase of transmission gaps (Fig.\ref{fig:Fig11r}c) and consequently the decrease of reliability (Fig.\ref{fig:Fig11r}a). As the $\rm B_{exp}$ parameter tends to 0, the scheduling algorithm drains the battery of the nodes and enforce their battery levels close to 0. This leads to the increase of transmission gaps (Fig.\ref{fig:Fig11r}f) and consequently the decrease of reliability (Fig.\ref{fig:Fig11r}d). As the $\rm B_{exp}$ tends to $\rm B_{max}$, the system wastes more energy (Fig.\ref{fig:Fig11r}e) and the reliability drops (Fig.\ref{fig:Fig11r}d), because less nodes are being selected to maintain the battery level close to the high $\rm B_{exp}$. Based on Fig.\ref{fig:Fig11r}b, the best configuration for the evaluation system parameters is $100\leq\rm B_{exp}\leq300$ and $\rm rlb_{min}\leq0.995$ where 0.995 is the highest value of estimation reliability.

\subsection{Scalability Study}

\begin{table}
\centering
\caption{Average results of 80 sensor nodes (24 real and 56 synthesized) for 30 days ($\rm B_{exp}=150$ and $\rm rlb_{min} = 0.98$).}
\begin{tabular}{lcccl}
\hline
\multicolumn{1}{|c|}{Algorithm} & \multicolumn{1}{c|}{\begin{tabular}[c]{@{}c@{}}Reliability\\ (\%)\end{tabular}} & \multicolumn{1}{c|}{\begin{tabular}[c]{@{}c@{}}Wasted \\ Energy (kJ)\end{tabular}} & \multicolumn{1}{c|}{\begin{tabular}[c]{@{}c@{}}Transmission\\Gaps\end{tabular}}\\ \hline
\multicolumn{1}{|l|}{RG} & \multicolumn{1}{c|}{84.9} & \multicolumn{1}{c|}{50.3} & \multicolumn{1}{c|}{1428}\\ \hline
\multicolumn{1}{|l|}{EG40} & \multicolumn{1}{c|}{61.5} & \multicolumn{1}{c|}{153.7} & \multicolumn{1}{c|}{21}\\ \hline
\multicolumn{1}{|l|}{EG70} & \multicolumn{1}{c|}{80.1} & \multicolumn{1}{c|}{64.6} & \multicolumn{1}{c|}{725}\\ \hline
\multicolumn{1}{|l|}{RR40} & \multicolumn{1}{c|}{28.9} & \multicolumn{1}{c|}{154.6} & \multicolumn{1}{c|}{23}\\ \hline
\multicolumn{1}{|l|}{RR70} & \multicolumn{1}{c|}{90.1} & \multicolumn{1}{c|}{65.4} & \multicolumn{1}{c|}{837}\\ \hline
\multicolumn{1}{|l|}{FDTS} & \multicolumn{1}{c|}{95.9} & \multicolumn{1}{c|}{52.8} & \multicolumn{1}{c|}{335}\\ \hline
\end{tabular}

\label{table:TableScRes}
\end{table}

Real water networks can cover a large area with a corresponding large number of nodes, which requires the scalability of the algorithm design.  In order to examine a larger scale smart water network, we exploit the raw data from the 24 sensor nodes and generate synthetic data of 56 virtual nodes, by virtually creating 2 neighbors for each real node (48 virtual nodes) and adding 8 additional virtual nodes in areas where the distance between two nodes was relatively large to the other nodes of the network. Then, the virtual nodes were distributed in equal distances between two real nodes. Hydraulic equations (i.e. equation(\ref{eq:equationl})), were used to generate the 30-day virtual pressure measurements. This process enable us to exploit the water network physics and avoids random data synthesis, resulting in realistic simulation studies.

The evaluation results of this scalability study are summarized in Table \ref{table:TableScRes}. RG algorithm waste less energy by 4.7\% than FAST-DTS, however it occurs 4.3 times more transmission gaps. EG40 and RR40 produce 314 less transmission gaps, nevertheless these algorithms are inappropriate for deployment because of the poor data reliability results (61.5\% and 28.9\% respectively). Overall, FAST-DTS managed to maintain the battery level by keeping energy waste and transmission gaps low, and it achieved the highest data reliability.

\section{Conclusion}
\label{sec:conclusion}
In this paper, we develop a reliable and sustainable wireless sensing system for high-sample-rate water pressure monitoring and abnormal behavior detection in smart water networks, powered by an energy harvesting system based on water flow.
In order to achieve sustainable sensor node operation by utilizing the dynamic and limited harvested energy, we use lossless compression and propose an in-node anomaly detection algorithm to reduce the raw data volumes (and therefore the energy) required for the high-power and long-range wireless transmission.
To further reduce energy consumption, our system only requires the transmission of a subset of the raw data streams, while other data streams are estimated using auto-regressive models instead of complex hydraulic models. The estimation is based on strong correlations among sensor data streams caused by the sound-velocity propagation of pressure signals inside the water network.
We formalize a stochastic optimization problem for the best selection of raw data transmissions that aim to maximize the aggregated estimation reliabilities, while guaranteeing a minimal reliability constraint and the sustainable operation of the smart water sensing system.
We develop DTS, a theoretically-proven asymptotically optimal solution to the formalized problem, based on Lyapunov optimization theory. Guided by the principles of DTS, we then propose FAST-DTS, a lightweight online algorithm that can adapt to arbitrary energy and correlation dynamics.

We evaluate our approach by using 170-day high sample rate data (128 samples per second) from our real smart water system and we compare our scheduling algorithm with three other algorithms. FAST-DTS outperforms these algorithms in terms of data reliability and sustainable operation, by achieving 99\% estimation accuracy and at the same time the lowest energy waste and transmission gaps. Currently, we are working to extend our approach by examining the impact of data estimation with regard to the automatic control process.

\appendix
\section*{APPENDIX}
\vspace{0.5em}

\begin{proof}[of Sustainable Operation]

Consider a scheduling decision $\mathcal{Y}(t)$ and an arbitrary node $j\in\mathcal{S}_b(t),~j\notin\mathcal{Y}(t)$. Let $\mathcal{Y}'(t)=\mathcal{Y}(t)\cup\{j\}$.
Let $\varphi_j(t)$ be the difference of (\ref{eq:perslotpiobj}) when scheduling decisions $\mathcal{Y}(t)$ and $\mathcal{Y}'(t)$ are adopted, we have
 \begin{eqnarray*}
 \varphi_j(t)&=&V\sum _{i\in\mathcal{S}}({U}_i(rlb_i(\mathcal{Y}(t)))+\sum _{i\in\mathcal{Y}(t)}a_i(t) -(V\sum _{i\in\mathcal{S}}({U}_i(rlb_i(\mathcal{Y'}(t)))+\sum _{i\in\mathcal{Y}'(t)}a_i(t)) \\
&=&V\sum _{i\in\mathcal{S}}({U}_i(rlb_i(\mathcal{Y}(t))) -{U}_i(rlb_i(\mathcal{Y'}(t))))-a_j(t)\\
 \end{eqnarray*}
 In order to ensure $ \varphi_j(t)>0$ when $B_j(t)<{\rm B_{exp}}$,
 we have $\forall E^{tr}_j(t),rlb_i(\mathcal{Y}(t)),rlb_i(\mathcal{Y}'(t))$
  \begin{equation*}
V\sum _{i\in\mathcal{S}}({U}_i(rlb_i(\mathcal{Y}(t))) -{U}_i(rlb_i(\mathcal{Y'}(t))))
>E^{tr}_j(t)(B_j(t)-{\rm B_{exp}}),\\
 \end{equation*}

Consider the facts that $0\leq rlb_i()\leq 1$, $E^{tr}_j(t)\leq E_{max}$, and $U_i()$ is an non-decreasing function, above inequality will always holds when $V<\frac{\sum _{i\in\mathcal{S}}(U_i(1)-U_i(0))}{\rm E_{max}B_{exp}}$.
\end{proof}

\begin{proof}[of Asymptotic Optimality]

To prove the optimality of DTS, we divide the system time horizon, $[1, {\rm t_{end}}]$, into $K$ successive frames with size $T$ intervals (i.e. $ {\rm t_{end}}=KT$). We assume that there exists an \textit{ideal} algorithm operating at the first interval of each frame $t=(k-1)T+1, 1\leq k\leq K$, which can obtain full information regarding the dynamics of the smart water system for the future $T$ slots (which is impossible in practice). Based on future knowledge, the ideal algorithm solves the following problem:

\begin{eqnarray}
&& \underset{\mathcal{Y}(t)}{\textbf{max}} \qquad\qquad\frac{1}{T} \sum_{t=kT-T+1}^{kT} \sum _{i\in\mathcal{S}}{U}_i(rlb_i(\mathcal{Y}(t)))
 \qquad\qquad \label{eq:opiobjideal}\\
 &&{\rm \textbf{subject~to}}\nonumber\\
&& \qquad\qquad\qquad\text{Constraints} (\ref{eq:musttransmitting})-(\ref{eq:ENOconstraint}) \nonumber\\
&&\frac{1}{T}\sum_{t=kT-T+1}^{kT}(h_i(t)-y_{i}(t) E^{tr}_i(t)- E^{in}_i(t))\geq0,\forall i\qquad\label{eq:congauranteeideal}
\end{eqnarray}

Let $\Phi(t)=\sum _{i\in\mathcal{S}}{U}_i(rlb_i(\mathcal{Y}(t)))$ achieved by the DTS algorithm, and $\Phi^{ideal}(k,T)$ denote the utility achieved by the ideal algorithm over each frame $1\leq k\leq K$.
We aim to prove the following inequality
\begin{eqnarray}
\frac{1}{\rm t_{end}}\sum_{1}^{\rm t_{end}}\Phi(t)\geq\frac{1}{K}\sum_{k=1}^{K}\Phi^{ideal}(k,T)-\frac{MT-N}{V}
\label{eq:theorem1result}
\end{eqnarray}
where $M=\frac{1}{2}|\mathcal{S}|\max({\rm h^2_{max}, {E'}^2_{max}})$ and $N=\frac{1}{\rm2 t_{end}}|\mathcal{S}|{\rm B^2_{max}}$.
 is a constant value and $\rm {E'}^2_{max}$ represents the upper bound energy costs for transmission and in-node operation for each interval, i.e. $E_{i}^{tr}(t)+E_{i}^{in}(t)\leq{\rm {E'}^2_{max}},\forall i,\forall t$ .

Inequality (\ref{eq:theorem1result}) shows that parameter $V$ can be set as large as desired to force $MT/V$ to be arbitrarily small. Specifically, Inequality (\ref{eq:theorem1result}) also demonstrates that when $T={\rm t_{end}}$, the optimal average aggregated utilities of reliability can be asymptotically achieved by DTS, as $V\rightarrow \infty$.
Now we prove the Inequality (\ref{eq:theorem1result}). Let $\triangle B_i(t)=B_i(t+1)-B_i(t)$ and define the Lyapunov function
\begin{equation}
L(t)=\frac{1}{2}\sum_{i\in\mathcal{S}}(B_i(t)-{\rm B_{exp}})^2
\end{equation}
Consider its one-interval drift plus penalty
 \begin{eqnarray}
\triangle _1L(t)&=&L(t+1)-L(t)-\Phi(t)\nonumber\\
&=& \frac{1}{2}\sum_{i\in\mathcal{S}}((B_i(t+1)-{\rm B_{exp}})^2-(B_i(t)-{\rm B_{exp}})^2)-\Phi(t)\qquad\nonumber\\
&=&\frac{1}{2}\sum_{i\in\mathcal{S}}(\triangle B_i^2(t)+2\triangle B_i(t)(B_i(t)-{\rm B_{exp}}))-\Phi(t)\nonumber\\
&\leq_{a}& M-V\Phi(t)-\sum_{i\in\mathcal{S}}(B_i(t)-{\rm B_{exp}})(y_{i}(t) E^{tr}_i(t)+ E^{in}_i(t)-h_i(t)))\label{eq:1-slotdrift}
 \end{eqnarray}
Inequality (\ref{eq:1-slotdrift}) is because of the fact that
  \begin{equation*}
M=\frac{1}{2}|\mathcal{S}|\max({\rm h^2_{max}, {E'}^2_{max}})\geq\frac{1}{2}\sum_{i\in\mathcal{S}}\triangle B_i^2(t)
 \end{equation*}
It can be easily seen that the DTS algorithm greedily minimizes the right-hand-side of inequality (\ref{eq:1-slotdrift}) at every interval $t$. Now we consider the $T$-interval drift of the Lyapunov function
 plus $T$-interval penalty

 \begin{eqnarray}
&&\triangle_TL(t)-V\sum_{t=kT-T+1}^{kT}\Phi(t)\nonumber\\
&&=L(kT)-L(kT-T+1)-V\sum_{t=kT-T+1}^{kT}\Phi(t)\nonumber\\
&&=\sum_{t=kT-T+1}^{kT}(\triangle_1L(t)-\Phi(t))\nonumber\\
&&\leq_{a} MT+\sqrt{2M}\frac{T(T-1)}{2}\nonumber\\
&&\quad+\sum_{t=kT-T+1}^{kT}(\sum_{i\in \mathcal{S}}(B_i(t)-{\rm B_{exp}})(y_{i}(t) E^{tr}_i(t)+ E^{in}_i(t)-h_i(t))-V\Phi(t))\nonumber\\
&&\leq_{b} MT^2-VT\Phi^{ideal}(k,T)\nonumber\\
&&\quad+\sum_{t=kT-T+1}^{kT}\sum_{i\in \mathcal{S}}(B_i(t)-{\rm B_{exp}})(y^{ideal}_{i}(t) E^{tr}_i(t)+ E^{in}_i(t)-h_i(t))
\label{eq:Tslotdriftpenalty}
\end{eqnarray}
where the inequality $\leq_a$ is based on inequality (\ref{eq:1-slotdrift}), the sum of $\triangle_1L(t)-V\Phi(t)$ over $T$ slots of the $k$th frame, and the fact that each queue backlog does not change by more than $(t-(kT-T+1))\max({\rm h_{max}, {E'}_{max}})$ for any slot $kT-T+1\leq t\leq kT$; the inequality $\leq_b$ follows from $M\geq\sqrt{M}, \forall M\geq 1$, and the fact that our DTS algorithm minimizes the right-hand side of the inequality $\leq_{a}$ over all possible transmission scheduling decisions, including the decisions of the ideal algorithm $y^{ideal}_i(t),i\in\mathcal{S}$, which achieves the optimal reliability utility of the ideal algorithm $\Phi^{ideal}(k,T)$. Consider (\ref{eq:Tslotdriftpenalty}) and the fact that $y^{ideal}_{i}(t)$ satisfies constraint (\ref{eq:congauranteeideal}), we have

\begin{equation}
\triangle_TL(t)-V\sum_{t=kT-T+1}^{kT}\Phi(t)\leq MT^2-VT\Phi^{ideal}(k,T) \label{eq:Tslotdriftpenaltyconcise}
\end{equation}

Taking a telescopic sum of the inequality (\ref{eq:Tslotdriftpenaltyconcise}) over $k\in\{1,...,K\}$ and dividing both side by $VKT$, we get
\begin{equation*}
\frac{L(KT+T)-L(1)}{KVT}-\frac{1}{KT}\sum_{t=1}^{KT}\Phi(t)\leq \frac{MT}{V}-\frac{1}{K}\sum_{k=1}^{K}\Phi^{ideal}(k,T)\qquad
\end{equation*}

Consider $\frac{L(KT+T)-L(1)}{KT}\geq N=\frac{1}{\rm2 t_{end}}|\mathcal{S}|{\rm B^2_{max}}$, we have
\begin{equation}
\frac{1}{KT}\sum^{KT}_{t=1}\Phi(t)\geq\frac{1}{K}\sum_{k=1}^{K}\Phi^{ideal}(k,T)-\frac{MT-N}{V}\nonumber\qquad\qquad\qquad\qquad
\end{equation}
Since ${\rm t_{end}}=KT$, Inequality (\ref{eq:theorem1result}) obviously holds.
\end{proof}

\end{document}